\documentclass[aps,preprint,superscriptaddress,groupedaddress]{revtex4-2}  
\usepackage{url}
\usepackage{breakurl}
\makeatletter
\g@addto@macro{\UrlBreaks}{\UrlOrds}
\g@addto@macro{\UrlBreaks}{\do\/\do\d}
\makeatother
\usepackage{xcolor}
\usepackage{graphicx}  
\usepackage{dcolumn}   
\usepackage{bm}        
\usepackage{amssymb,amsfonts,amsmath}   
\usepackage{tabu}
\usepackage{tabularx}
\usepackage{textgreek}
\usepackage{gensymb}
\usepackage{textcomp}
\usepackage{float}
\usepackage{placeins}
\usepackage[version=4]{mhchem}
\usepackage{xr}
\makeatletter
\newcommand*{\addFileDependency}[1]{
	\typeout{(#1)}
	\@addtofilelist{#1}
	\IfFileExists{#1}{}{\typeout{No file #1.}}
}\makeatother

\newcommand*{\myexternaldocument}[1]{%
	\externaldocument{#1}%
	\addFileDependency{#1.tex}%
	\addFileDependency{#1.aux}%
}
\myexternaldocument{\detokenize{Symmetry_Perturbation_PRSA_SI_v3}}

\graphicspath{{/home/alexander/Dropbox/Symmetry/Approx symmetry/figures/}}

\begin{document}
	
	\title{Approximate Lie symmetries and singular perturbation theory}

	\author{Alexander J. Dear}
	\affiliation{School of Engineering and Applied Sciences, Harvard University, USA}
	\affiliation{Center for Molecular Protein Science, Lund University, 221 00 Lund, Sweden}
	\affiliation{To whom correspondence should be addressed. E-mail: adear@g.harvard.edu}
	
	\author{L. Mahadevan}
	\affiliation{School of Engineering and Applied Sciences, Departments of Physics and Organismic and Evolutionary Biology, Harvard University, USA}
	\affiliation{To whom correspondence should be addressed. E-mail: lmahadev@g.harvard.edu}

	\date{\today}

\begin{abstract}
Singular perturbation theory plays a central role in the approximate solution of nonlinear differential equations. However, applying these methods is a subtle art owing to the lack of globally applicable algorithms. Inspired by the fact that all exact solutions of differential equations are consequences of (Lie) symmetries, we reformulate perturbation theory for differential equations in terms of expansions of the Lie symmetries of the solutions. This is a change in perspective from the usual method of obtaining series expansions of the solutions themselves. We show that these approximate symmetries are straightforward to calculate and are never singular; their integration is therefore an easier way of constructing uniformly convergent solutions. This geometric viewpoint naturally subsumes the RG-inspired approach of Chen, Goldenfeld and Oono, the method of multiple scales, and the Poincare-Lindstedt method, by exploiting a fundamental class of symmetries that we term ``hidden scale symmetries''. It also clarifies when and why these singular perturbation methods succeed and just as importantly, when they fail. More broadly, direct, algorithmic identification and integration of these hidden scale symmetries permits solution of problems where other methods are impractical.
\end{abstract}

\maketitle

\section{Introduction}
How do we solve differential equations (DEs) analytically? We guess a solution, or use various approximate methods that fall under the rubric of perturbation theory. The latter involve a whole range of methods but it is never clear which will work for which equation. So, this would appear to be a laborious process, with uncertain chances of success. However, it is known that all exact solutions of arbitrary DEs, should they exist can be approached via an algorithmic manner using the theory of continuous (Lie) symmetry groups~\cite{Olver2000}. (Indeed, even though today it finds application across diverse fields of the physical sciences, solving DEs was the original motivation for the development of Lie theory.) Lie theory reveals that in the  cases where an exact solution to a DE can be found, the underlying property that is unconsciously or explicitly exploited is a Lie symmetry of the DE~\cite{Stephani1990}. Lie theory furthermore provides a single general integration procedure that can be applied to any arbitrary DE, in which its Lie symmetries are first calculated algorithmically, and then leveraged to simplify or solve the DE using a small number of standard techniques.

Many differential equations (DEs) encountered in the physical sciences do not admit simple exact analytical solutions, regardless of the method used. Therefore, solving DEs approximately using perturbation theory is an important tool. It involves identifying a small or switching parameter, $\epsilon$, in the DEs such that they admit a simple analytical solution for $\epsilon=0$. The dependent variable $y$ is then solved for as a ``perturbation series'' in $\epsilon$, $y^{(0)}+\epsilon y^{(1)}+\dots$~\cite{Bender1999I}. A critical limitation of perturbation theory is that many problems are ``singular'', defined as one whose perturbation series has a radius of convergence that vanishes somewhere within the region of phase space of interest~\cite{Bender1999I}.  To overcome this issue, there exist several ``singular perturbation methods'' that convert divergent perturbation series into uniformly convergent approximate solutions~\cite{Bender1999I,Kevorkian1996,Hinch1991}. Their implementation requires considerable intuition and thus, analogously to exact solution methods, it is challenging to determine \textit{a priori} to which kinds of divergent series a given technique can be successfully applied.

This prompts us to ask the question: just as (exact) Lie symmetries underpin most methods for the exact solution of DEs, could \textit{approximate} Lie symmetries (appropriately defined) provide a rigorous and unified basis for the various singular perturbation methods used to approximately solve perturbed DEs? Equivalently, we ask if there is a natural approximate scheme inspired by the exact Lie-group inspired framework in quantum field theory~\cite{bogoliubov} that led to modern renormalization group based techniques (for a historical review, see~\cite{shirkov}). From the perspective of this study that focuses on classical mechanics, the renormalization group-inspired method of Chen, Goldenfeld and Oono (CGO RG)~\cite{Chen1994,Chen1996} and its alternative formulation by Kunihiro (RG/E)~\cite{Kunihiro1995,Kunihiro1997}, have been shown to equal or exceed in accuracy and generality earlier methods for regularizing singular perturbation series, being simpler than matched asymptotic expansions, and an often faster route to finding the slow motion equations sought by multiple scales analysis and reductive perturbation techniques~\cite{Lagerstrom1972, Hinch1991}. Discovering a Lie symmetry basis for these techniques would help rationalize their success, and could be used to understand more clearly under what circumstances they will be inapplicable.

\section{Structure of paper}

For completeness, we provide in Sec.~\ref{sec:methods:Lie} a quick introduction to those aspects of the established Lie theory of differential equations that are needed in this paper. In the Supporting Information (SI) Sec.~S1 we also provide overviews of singular perturbation theory, the CGO RG method, and its alternative RG/E formulation.

Our first result is the establishment of an algorithmic integration procedure for perturbed DEs based on approximate Lie symmetries, that always yields convergent solutions (Sec.~\ref{sec:Lieintegration}). Next, in Sec.~\ref{sec:symcalc} we develop a simple and practical method for calculating approximate Lie symmetries of solutions to perturbed DEs, illustrating its application using the simple linear example of the overdamped linear oscillator. In Sec.~\ref{sec:hiddenscale} we devise a conceptually transparent and highly general singular perturbation method based on exploiting a class of symmetries that we term ``hidden scale'' symmetries. We showcase some of its advantages over other singular perturbation techniques in Sec.~\ref{sec:highorder} by solving a difficult fourth-order Turing-type pattern formation equation in a relatively straightforward manner. 

In Sec.~\ref{sec:derivingCGO} we next \textit{derive} the CGO RG method, revealing it to simply be an indirect approach to identifying and exploiting hidden scale symmetries involving some trial-and-error. The celebrated solutions to first order using CGO RG of the notorious ``switchback'' problems, that would appear to contradict this Lie symmetry interpretation, are revisited in Sec.~\ref{sec:switchback} and found to actually be \textit{regular} perturbation series in a switching parameter entering the boundary conditions. The Oseen equation of fluid mechanics is found in Sec.~\ref{sec:Oseen} to be the regular first-order boundary condition perturbation of the steady incompressible Navier-Stokes equation, explaining its successful use to approximate the latter. The analytical regular perturbative solution for flow past a cylinder in the low-Reynolds limit is thus derived.

We follow with a series of examples to which we apply the hidden scale symmetry method to further demonstrate its advantages and to reveal the basis in Lie symmetries of other popular techniques. First, we revisit the ``switchback'' problems in Sec.~\ref{sec:exactsym}, this time to second order, to clarify using its Lie symmetry basis precisely why CGO RG sometimes finds exact solutions. Next, in Sec.~\ref{sec:MMS} we use the Mathieu equation as an example that is solved by the method of multiple scales (MMS). Solving directly by hidden scales symmetry reveals that MMS also seeks these hidden scale symmetries, but indirectly and often involving significant guesswork and/or additional effort. Following this, we explain in Sec.~\ref{sec:PLM} the hidden scale symmetry basis of the Poincare-Lindstedt method (PLM), and use this to determine strict guidelines around its applicability. We use the Korteweg-de-Vries equation as an example that has been erroneously treated with PLM in the literature, resulting in incorrect solutions.

Finally, we use a modified Burgers equation to show in Sec.~\ref{sec:pertsym} that when these hidden scale symmetry-based methods fail, other classes of approximate symmetries can sometimes be similarly exploited to develop global solutions. We conclude in Sec.~\ref{sec:conclusions} with a discussion of several thought-provoking questions raised by the ideas introduced in this paper.

\section{Review of Lie point symmetries of differential equations}\label{sec:methods:Lie}

A point transformation is one that transforms independent variables $x$ and dependent variables $y$ of the object being acted upon, to $\tilde{x}$ and $\tilde{y}$. In the context of differential equations other kinds of transformations are possible (namely, those acting upon derivatives and integrals of the dependent variables), but we can ignore these here. Point transformations that are indexed by at least one arbitrary parameter $s$ may be written $\tilde{x}=\tilde{x}(x,y,s),\ \tilde{y}=\tilde{x}(x,y,s)$. When these are also invertible, contain the identity at $s=0$, and obey $\tilde{x}(\tilde{x}(x,y,s),\tilde{y}(x,y,s),t)=\tilde{x}(x,y,s+t)$, they form a one-parameter (or multi-parameter) group of point transformations.

\begin{figure}
	\includegraphics[width=\textwidth]{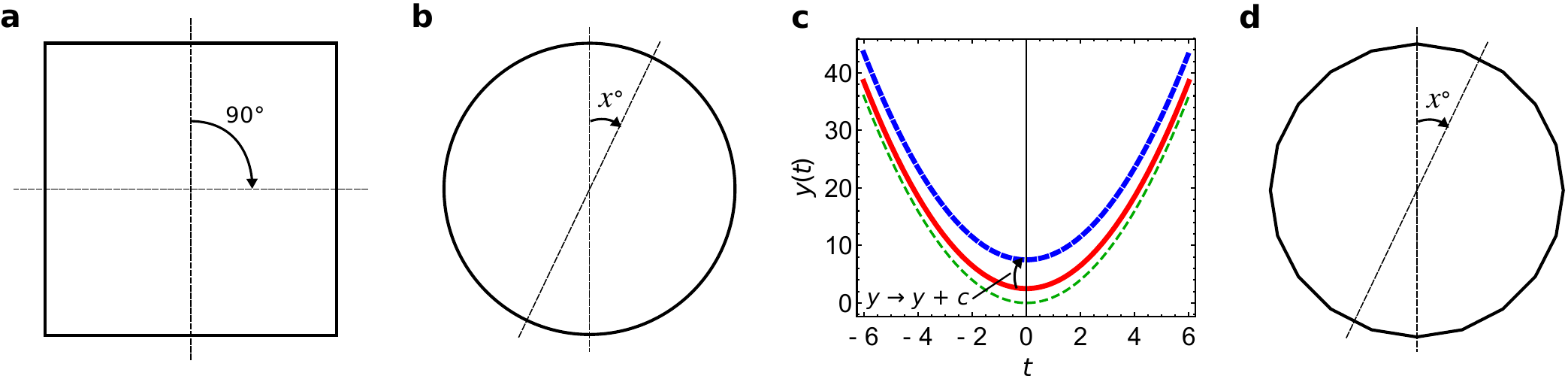}
	\caption{An overview of Lie symmetries. \textbf{a}: Squares have discrete rotational symmetries. These cannot be reduced to infinitesimal form; therefore, they are not Lie symmetries. \textbf{b}: Circles can be rotated by any amount; rotation is thus a Lie symmetry of the circle. \textbf{c}: In general, symmetries of DEs map solutions to other solutions with different boundary conditions. An arbitrary translation on the $y$ axis is a Lie symmetry of the DE $\dot{y}=2t$, because this is solved by $y=t^2+c$, and the translation just changes the value of $c$, giving the solution to the DE for new boundary conditions. \textbf{d}: Dodecagons are only approximately invariant under infinitesimal rotational transformations (to $O(\varepsilon)$, where $\varepsilon\sim z\cos\theta$, with $\theta$ the external angle and $z$ the side length), which are therefore an approximate Lie symmetry.}
	\label{Fig1}
\end{figure}

Such transformations have the useful property that they are continuous, and thus we can consider the infinitesimal transformation by expanding around $s=0$:
\begin{subequations}
	\begin{align}
	\tilde{x}(x,y,s)&=x+s\frac{\partial\tilde{x}}{\partial s}\bigg|_{s=0}+\dots=x+s\bm{X}x+O(s^2)\\
	\tilde{y}(x,y,s)&=y+s\frac{\partial\tilde{y}}{\partial s}\bigg|_{s=0}+\dots=y+s\bm{X}y+O(s^2),
	\end{align}
\end{subequations}
where the operator $\bm{X}$ is:
\begin{equation}
\bm{X}=\xi(x,y)\frac{\partial}{\partial x}+\eta(x,y)\frac{\partial}{\partial y},
\end{equation}
and the elements of the tangent vector $(\xi(x,y),\eta(x,y))$ are:
\begin{equation}
\xi(x,y)=\frac{\partial\tilde{x}}{\partial s}\bigg|_{s=0},\quad\eta(x,y)=\frac{\partial\tilde{y}}{\partial s}\bigg|_{s=0}.
\end{equation}
The operator $\bm{X}$ is the infinitesimal generator of the point transformation. Integrating the tangent vector over $s$ will yield a finite transformation. Now we can define a Lie point symmetry of an object as a continuous point transformation that leaves the object invariant (Fig.~\ref{Fig1}\textbf{a}-\textbf{b}). Other kinds of Lie symmetry exist but we need not consider them here.

Lie symmetries of $F=0$, where they exist, are found by solving:
\begin{equation}\label{DEsym}
\bm{X}F|_{F=0}=0.
\end{equation}
This is an essentially algorithmic procedure but is often extremely laborious. Nowadays it is therefore usually automated using packages available for computer algebra systems (CAS) such as Mathematica. We omit the details of this procedure because solving Eq.~\eqref{DEsym} is not necessary in the present work.

Similarly, only one of the several established methods for integrating or simplifying DEs using their symmetries is relevant enough to this paper to merit a description here. Lie symmetries of differential equations transform solutions to other solutions (Fig.~\ref{Fig1}\textbf{c}). Thus, if a special solution for a partial differential equation is known, families of group-invariant solutions may be found by identifying the Lie symmetries of the parent PDE and applying the corresponding finite transformations to this special solution. As a very brief illustration of this, it may be easily found by CAS that the heat equation $u_t-u_{xx}=0$ admits several point symmetries, one of which is:
\begin{equation}
\bm{X}=xt\frac{\partial}{\partial x}+t^2\frac{\partial}{\partial t}-\left(\frac{x^2}{4}+\frac{t}{2}\right)u\frac{\partial}{\partial u}.
\end{equation}
Integrating an infinitesimal symmetry generator over a finite distance in parameter space leads to finite transformations, also known as orbits, motions along which leave the DE invariant. They can be computed by solving $\bm{X}J=0$ for general $J$ using the method of characteristics. For this particular symmetry, $J$ is constant along characteristics defined by:
\begin{equation}
\frac{d\tilde{x}}{ds}=\tilde{x}\tilde{t},\quad \frac{d\tilde{t}}{ds}=\tilde{t}^2,\quad \frac{d\tilde{u}}{ds}=-\left(\frac{\tilde{x}^2}{4}+\frac{\tilde{t}}{2}\right)\tilde{u},
\end{equation}
(where $s$ can be interpreted as the group parameter). Integrating from $(s,\tilde{t},\tilde{x},\tilde{u})=(0,t,x,u)$ yields:
\begin{equation}
	\tilde{x}=x/(1-s t),\quad\tilde{t}=t/(1-s t),\quad\tilde{u}=u\sqrt{1-s t}e^{-sx^2/4(1-s t)}.
\end{equation}
They may be inverted and combined to give:
\begin{equation}
\tilde{u}=\frac{u}{\sqrt{1+s \tilde{t}}}e^{-s\tilde{x}^2/4(1+s \tilde{t})}.
\end{equation}
Applying this to the special solution $u=A$, where $A$ is an arbitrary constant, yields the group-invariant solution (dropping the tildes):
\begin{equation}
u=\frac{A}{\sqrt{1+s t}}e^{-sx^2/4(1+s t)},
\end{equation}
and $s$ is now another arbitrary constant.

Finally, most of the established literature deals with exact symmetries of DEs, whose transformations leave them entirely unchanged. A more recent insight~\cite{Baikov1988} is that the symmetry generator of a perturbed DE can be expanded in powers of $\epsilon$. Truncating this series yields ``approximate symmetries'', that leave the DE invariant up to a given order in $\epsilon$ (Fig.~\ref{Fig1}\textbf{d}). They can be identified by solving:
\begin{equation}\label{appDEsymfind}
(\bm{X}^{(0)}+\epsilon\bm{X}^{(1)}+...)(F_0+\epsilon F_1)|_{F_0+\epsilon F_1=0}=0, 
\end{equation}
order-by-order~\cite{Ibragimov2009}. Systems that do not admit exact solutions may admit such approximate symmetries, which can then be used to integrate approximately the DEs to yield approximate solutions.\footnote{Note alternative approaches for their computation have been proposed~\cite{Fushchich1989,Pakdemirli2004}, and there is some debate as to which is superior~\cite{Pakdemirli2004,Wiltshire2006}.} However, they are more technically challenging to calculate than exact symmetries, and few or no CAS packages can fully automate this procedure.

\section{Results}

\subsection{Approximate Lie symmetry is a natural language for perturbation theory}\label{sec:Lieintegration} 

\begin{figure*}
	\centering
	\includegraphics[width=0.85\textwidth]{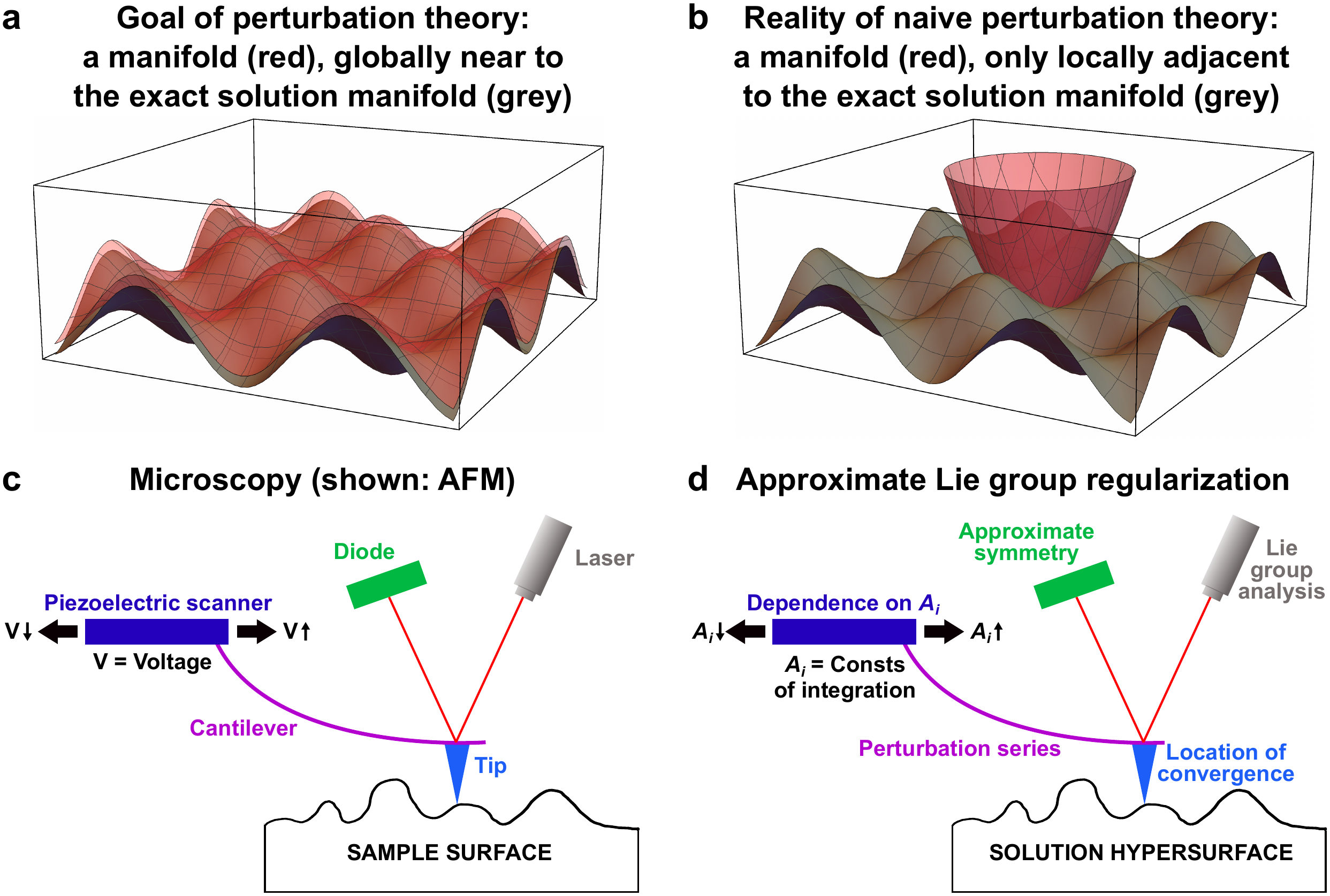}
	\caption{\textbf{a}-\textbf{b}: Singular perturbation theory can be thought of in geometric terms, as an attempt to approximate the exact solution manifold (grey). \textbf{a}: the goal is to construct a simple analytical solution whose manifold (red) is everywhere within $\epsilon$ of the exact manifold. \textbf{b}: for singularly perturbed problems, a naive application of perturbation theory generates a manifold that coincides only locally with the exact manifold. \textbf{c}-\textbf{d}: Analogy between our approximate Lie group regularization approach and microscopy. (Although we illustrate the analogy using atomic force microscopy (AFM), it also works for other kinds of microscopy such as optical microscopy.) \textbf{c}: In AFM, the sample surface is scanned by the tip of a probe by changing the voltage of a piezoelectric scanner connected to the probe. The tip displacement is determined by detecting the deflection of a laser beam by the cantilever using a diode. A model of the surface can thus be built, whose resolution is given by the radius of the probe tip. \textbf{d}: In our approximate Lie group approach, the solution hypersurface is scanned by the convergent region of a singular perturbation series in $\varepsilon$ by changing the values of the constants of integration $A_i$. The position in phase space of the point of convergence can be monitored during this process by approximate Lie group analysis and recorded in an approximate symmetry. An approximate solution can then be built from this symmetry, whose accuracy is given by the magnitude of $\varepsilon$.}
	\label{Fig2}
\end{figure*}

In geometric terms, the fundamental aim of singular perturbation theory is to construct a manifold that approximates everywhere within the phase space of interest to $O(\epsilon^n)$ the manifold spanned by the exact solution to a perturbed DE with perturbation parameter $\epsilon$ (Fig.~\ref{Fig2}\textbf{a}). Perturbation theory in the traditional sense (expanding the dependent variable as a power series in $\epsilon$) only finds such approximate manifolds if this series happens to be regular. Otherwise, it finds a manifold that coincides only locally with an approximate manifold (Fig.~\ref{Fig2}\textbf{b}). By contrast, $O(\epsilon^n)$ expansions of the \textit{Lie symmetries} of the solution to a perturbed DE (approximate Lie symmetries~\cite{Baikov1988,Ibragimov2009}) must encode transformations that map both exact and approximate manifolds to themselves to $O(\epsilon^n)$ \textit{globally}. A specific $O(\epsilon^n)$ approximate manifold is moreover left exactly invariant by a given $O(\epsilon^n)$ approximate symmetry.

This gives us the idea of constructing globally valid approximate solutions to a perturbed DE by explicitly identifying and exploiting an approximate solution symmetry. The orbits of such symmetries describe pathways running along the associated approximate manifold, obtained by repeated transformation of a given starting point on the manifold. So, transforming with this symmetry a special solution valid at one value of this coordinate should yield a global approximate solution. The power of this approach can be greatly increased by treating nominally constant parameters in the problem as independent variables, and looking for approximate solution symmetries that act on them. Not only does this increase the likelihood of a suitable symmetry existing (that we term approximate ``\textit{extended} symmetries''), but it also makes easier the identification of a suitable special solution to transform, which can now be e.g.\ a solution valid when one of the constant parameters is set to zero.

\subsection{Approximate solution symmetries can be calculated directly from the perturbation series}\label{sec:symcalc} 
The solutions to a DE admit symmetries that the DE itself does not, and vice versa. The computation of solution symmetries, exact or approximate, has hitherto been done by applying standard techniques of Lie analysis to the DEs themselves, either with auxiliary ``differential constraints''~\cite{Olver1986}, or with a subsequent ``restriction'' step to eliminate unwanted degrees of freedom associated with non-fixed initial or boundary conditions~\cite{Kovalev1998}. Unfortunately both approaches can be extremely computationally challenging, even with the aid of CAS, particularly for approximate extended symmetries; moreover, neither approach can reliably find all solution symmetries.

Fortunately, if a perturbation series solution with arbitrary integration constants is available (a ``bare'' perturbation series), it proves possible to compute \textit{approximate} solution symmetries in a much more direct manner, with minimal calculations required. The manifold of a singular perturbation series calculated to $O(\epsilon^n)$ locally approximates the exact solution manifold to $O(\epsilon^n)$ near where the boundary or initial conditions are imposed, so the series must still locally be left invariant by the $O(\epsilon^n)$ approximate solution symmetries of the perturbed DE. But if the integration constants $A_i$ are arbitrary then these conditions can be imposed anywhere. So, by changing the values of these constants appropriately, the perturbation series can be made to act like a probe, locally satisfying the approximate solution symmetries at any desired region of the manifold (analogously to how microscopy works, Fig.~\ref{Fig2}\textbf{c}-\textbf{d}).

So, instead of employing all the complicated mathematical machinery necessary to calculate the approximate symmetries of the parent differential equation (Eq.~\eqref{appDEsymfind}), we can merely calculate the symmetries of the perturbation series, which must also be approximate symmetries of the exact solution. This is achieved by solving to order-by-order the equation:
\begin{equation}\label{approxsym}
(\bm{X}^{(0)}+\epsilon\bm{X}^{(1)}+\dots)(y-y^{(0)}-\epsilon y^{(1)}-\dots)|_{y=y^{(0)}+\epsilon y^{(1)}+\dots}=0,
\end{equation}
where $y^{(i)}$ is the $i$\textsuperscript{th} term in the naive perturbation series solution (for arbitrary initial or boundary conditions) to the differential equation $F(y)=0$.	

The much greater simplicity of this approach opens up new possibilities for Lie symmetry analysis. A key such possibility is that this method can identify extended symmetries that act not just on the parameters entering the DEs, but also on parameters such as the integration constants $A_i$ that enter solely the solutions. There is no way to do this by the traditional approach of identifying the symmetries of the parent differential equation. In particular, an approximate symmetry describing how the integration constants $A_i$ change with $x$ should in principle provide a global approximation of the exact solution anywhere on the manifold. To explore this hypothesis, and to demonstrate how Eq.~\eqref{approxsym} is solved in practice, we consider the overdamped harmonic oscillator for $t\geq 0$ and undefined initial conditions:
\begin{equation}
\epsilon\frac{d^2y}{dt^2}+\frac{dy}{dt}+y=0,\quad \epsilon\ll 1.
\end{equation}
Transforming into the inner layer by writing $\tau=\epsilon t$, this becomes:
\begin{equation}\label{overdampedHO}
\frac{d^2y}{d\tau^2}+\frac{dy}{d\tau}+\epsilon y=0.
\end{equation}
Although it may be solved exactly, it is also a singular perturbation problem (Fig.~\ref{Fig3}\textbf{a}). Expanding $y$ in $\epsilon$ as $y(\tau,\epsilon)=y^{(0)}(\tau)+\epsilon y^{(1)}(\tau)+O(\epsilon^2)$ yields the following perturbation equations to first order:
\begin{equation}\label{overdampedperteqs}
\frac{d^2y^{(0)}}{d\tau^2}+\frac{dy^{(0)}}{d\tau}=0,\quad \frac{d^2y^{(1)}}{d\tau^2}+\frac{dy^{(1)}}{d\tau}+ y^{(0)}=0.
\end{equation}
Solving without initial conditions yields the ``bare'' perturbation series:
\begin{equation}\label{Eq33}
y_1(\tau)=A+Be^{-\tau}+\epsilon\left[-A\tau +B\tau e^{-\tau}\right],
\end{equation}
where the subscript indicates a finite perturbation series $y_n=\sum_{i=0}^{n}\epsilon^iy^{(i)}$. The generator to first order in $\epsilon$ for a symmetry connecting $A$, $B$ and $\tau$ has the form:
\begin{equation}\label{X1}
\bm{X}_1=\sum_{i=0}^1\epsilon^i\bm{X}^{(i)}=\sum_{i=0}^1\epsilon^i\left(\xi_A^{(i)}\frac{\partial}{\partial A}+\xi_B^{(i)}\frac{\partial}{\partial B}+\xi_{\tau}^{(i)}\frac{\partial}{\partial \tau}\right).
\end{equation}

It should now be clear that solving Eq.~\eqref{approxsym} alone using Eq.~\eqref{X1} will \textit{not} give a single symmetry that tells us how to update both $A$ and $B$ in response to an arbitrary change in $\tau$. This is because only one degree of freedom is eliminated at each order, leaving either $\xi_A^{(i)}$ or $\xi_B^{(i)}$ also arbitrary. This issue is resolved by noting that a globally valid approximate symmetry should also leave the derivatives of the solution approximately invariant. So, alongside Eq.~\eqref{approxsym}, we also require:
\begin{equation}\label{approxsymderiv}
\bm{X}_1\left(\dot{y}-\dot{y}_1\right)\big|_{y=y_1}=0.
\end{equation}
Solving Eqs.~\eqref{approxsym} and~\eqref{approxsymderiv} order-by-order in $\epsilon$, we find at zeroth order:
\begin{subequations}
	\begin{align}
	\bm{X}^{(0)}y^{(0)}&=\xi_A^{(0)}+\xi_B^{(0)}e^{-\tau}-Be^{-\tau}\xi_\tau^{(0)}=0\\
	\bm{X}^{(0)}\dot{y}^{(0)}&=-\xi_B^{(0)}e^{-\tau}+Be^{-\tau}\xi_\tau^{(0)}=0\\
	\therefore \xi_A^{(0)}&=0,\quad \xi_B^{(0)}=B\xi_\tau^{(0)}.
	\end{align}
\end{subequations}
Then, at first order, and seeking only zeroth-order transformations of $\tau$:
\begin{subequations}
	\begin{align}
	\bm{X}^{(0)}\dot{y}^{(1)}+\bm{X}^{(1)}\dot{y}^{(0)}&=e^{-\tau}\left(B\xi_\tau^{(1)}-\xi_B^{(1)}-B\xi_\tau^{(0)}\right)\\
	\bm{X}^{(0)}y^{(1)}+\bm{X}^{(1)}y^{(0)}&=-A\xi_\tau^{(0)} +B e^{-\tau}\xi_\tau^{(0)} +\xi_A^{(1)}+\xi_B^{(1)}e^{-\tau}-Be^{-\tau}\xi_\tau^{(1)}\\
	\therefore \xi_B^{(1)}&=-B\xi_\tau^{(0)},\quad \xi_A^{(1)}=A\xi_\tau^{(0)}.
	\end{align}
\end{subequations}
Pulling this together yields the generator of the anticipated $A_i$--$x$ symmetry:
\begin{equation}\label{symoverdamped}
\bm{X}_1=\xi_\tau^{(0)}\left(\frac{\partial}{\partial\tau}+\epsilon A\frac{\partial}{\partial A}+B(1-\epsilon)\frac{\partial}{\partial B}\right).
\end{equation}

Symmetries connecting integration constants and independent variables may seem unusual, but in fact arbitrary transformations of the integration constants are clearly always symmetries of DEs. In this context they are known as ``generalized symmetries''~\cite{Olver2000} or ``dynamical symmetries''~\cite{Stephani1990}, and are closely studied because they underpin Noether's theorem~\cite{Olver2000}. For symmetries instead of the solutions to DEs, the permissible transformations of the $A_i$ are no longer arbitrary, and we need new terminology. Since, as will become clear later, these symmetries find hidden scales, we refer to them hereafter as ``hidden scale symmetries''.

\subsection{General formulation of the ``hidden scale'' symmetry method}\label{sec:hiddenscale}

As $\tau\to 0$, the divergent terms of the perturbation series Eq.~\eqref{Eq33} vanish, yielding a non-divergent special solution $y=A+B$. Because the generator Eq.~\eqref{symoverdamped} is valid globally to $O(\epsilon)$, so must be its orbits, given by:
\begin{equation}
\frac{dA}{d\tau}=\epsilon A,\quad \frac{dB}{d\tau}=B(1-\epsilon).
\end{equation}
These orbits can therefore be used to transform the non-divergent $\tau=0$ solution into a global approximate solution. So, integrating backwards from $\tau=\tau$ to $\tau=0$ gives the finite transformation connecting $(A,\ B)$ at $\tau=0$ with their values $(\tilde{A},\ \tilde{B})$ at $\tau=\tau$:
\begin{equation}
A=\tilde{A} e^{-\epsilon\tau},\quad B=\tilde{B} e^{-(1-\epsilon)\tau},
\end{equation}
representing directly the time dependence of the constants of integration that is required for the perturbation series to be valid for all $\tau$. These orbits must be true to $O(\epsilon)$ for any $\tau$; substituting into the non-divergent $\tau=0$ solution $y=A+B$ thus provides a genuine global $O(\epsilon)$ perturbative solution to Eqs.~\eqref{overdampedHO}:
\begin{equation}\label{HSsymsoln}
y(\tau)=\tilde{A} e^{-\epsilon\tau}+\tilde{B} e^{-(1-\epsilon)\tau}+O(\epsilon^2).
\end{equation}

We will now derive a simplified but general formulation for this approach. Insodoing, we will show that identifying and solving the finite transformation (FT) equations for hidden scale symmetries like this can be performed directly, with no need to explicitly calculate the symmetries themselves.

The element of the tangent vector in the $x$ direction may be chosen without loss of generality to equal 1. For an arbitrary $n$\textsuperscript{th}-order ODE of the form:
\begin{equation}\label{arbitraryODE}
F\left(x,y_j,y_j',y_j'',\dots,y_j^{(n)}\right)=0,
\end{equation}
the generator $\bm{X}_k(x,A_i)$ of the $k$\textsuperscript{th}-order hidden scale symmetry may thus be written:
\begin{equation}\label{Dsym_general}
\bm{X}_k(x,A_i)=\frac{\partial}{\partial x}+\sum_{i=1}^n\sum_{j=0}^k\epsilon^j\xi_{A_i}^{(j)}\frac{\partial}{\partial A_i}.
\end{equation}
An invariant $J_k(x,A_i)$ of this approximate symmetry satisfies $\bm{X}_k(x,A_i)J_k(x,A_i)=0$. They are excellent approximate solutions because $J_k(x,A_i)=y(x,A_i)+O(\epsilon^{k+1})$ globally. The equation that gives this invariant $J_k(x,A_i)$ is:
\begin{equation}
0=\bm{X}_k(x,A_i)J_k=\frac{\partial J_k}{\partial x}+\sum_{i=1}^n\sum_{j=0}^k\epsilon^j\xi_{A_i}^{(j)}\frac{\partial J_k}{\partial A_i}.
\end{equation}
From the method of characteristics 
this partial differential equation is reduced to $dJ_k/dx=0$, i.e.\ $J_k=$ const., along characteristics defined by:
\begin{equation}
\sum_{j=0}^k\epsilon^j\xi_{A_i}^{(j)}=\frac{dA_i}{dx}.
\end{equation}
Now, using this result in Eq.~\eqref{Dsym_general} yields:
\begin{equation}
\bm{X}_k(x,A_i)=\frac{\partial }{\partial x}+\sum_i\frac{dA_i}{dx}\frac{\partial}{\partial A_i},
\end{equation}
where the $A_i$ are now functions of $x$ that span the $J_k(x,A_i)$ invariant (the orbits of the symmetry). If we have not already computed the hidden scale symmetry then $\xi_{A_i}^{(j)}$ and thus $dA_i/dx$ are not known. They may, however, be in principle calculated by observing that the perturbation series to order $k,\ y_k$, is left invariant by the hidden scale symmetry only in the limit $\epsilon\to 0$, i.e. $\bm{X}_k(x,A_i)y_k|_{\epsilon\to 0}=0$. Thus the finite transformation (FT) equation can be rewritten as:
\begin{equation}\label{DearRGeq}
\frac{\partial y_k}{\partial x}\bigg|_{\epsilon\to 0}+\sum_i\frac{dA_i}{dx}\frac{\partial y_k}{\partial A_i}\bigg|_{\epsilon\to 0}=0.
\end{equation}

Eq.~\eqref{DearRGeq} is underdetermined when there is more than one orbit to determine, i.e.\ more than one integration constant. This is resolved by requiring both the solution manifold and its first $n-1$ derivatives to be left approximately invariant. Since these can be approximated at any location by the equivalent derivative of the perturbation series in the limit $\epsilon\to 0$, we solve the following equations for $j=[0,\dots,n-1]$:
\begin{equation}\label{DearRGeq2}
\frac{\partial^j y_k}{\partial x^j}\bigg|_{\epsilon\to 0}+\sum_{i=1}^n\frac{dA_i}{dx}\frac{\partial}{\partial A_i}\frac{\partial^j y_k}{\partial x^j}\bigg|_{\epsilon\to 0}=0.
\end{equation}
This provides $n$ equations for $n$ unknowns, fully determining the system.

Only the terms divergent in $x$ need be removed to yield the desired approximate manifold. Therefore, it is mathematically acceptable, and sometimes convenient, to relabel the divergent instances of $x$ as a new variable $\mu$ (we call this ``painting'' the divergent terms, and discuss it in greater depth in SI Sec.~S2). We can then solve the FT Eqs.~\eqref{DearRGeq2} with respect to $\mu$ rather than $x$, demanding independence from $x$ and making the reverse substitution $\mu\to x$ in the final step, as will be illustrated later. One can also simplify further by deleting all bar the most divergent terms at each order prior to solving the FT equations, at the price of the resulting symmetry retaining full validity only asymptotically (also discussed later). Following these prescriptions, Eqs.~\eqref{DearRGeq2} directly finds the simplest possible $k$th-order perturbative approximation to the solution manifold; that is, the invariant of the $O(\epsilon^k)$ hidden scale symmetry, which constitutes a globally valid $O(\epsilon^k)$ approximate solution.

\subsection{Hidden scale symmetry method is especially well-suited for solving high-order ODEs}\label{sec:highorder}
Higher-order ODEs give rise to multiple constants of integration, higher-order divergent terms in the perturbation series, and frequently unusual timescales. This makes most other singular perturbation methods painstaking to implement, including CGO RG and MMS. Such problems are thus ideal for demonstrating the power and simplicity of the hidden-scale symmetry method.

We consider the formation of localized deformations in a flexible filament that can swell or grow in length, but that is embedded in an elastic substrate, studied in~\cite{Michaels2019c}. These are described by sinusoidal displacements perpendicular to the original alignment of the filament, modulated by a non-trivial envelope $\bar{W}$. Making the substitution $y=\bar{x}-\bar{L}/2$ in Eq.~(3.2) from~\cite{Michaels2019c} and restoring a term that was dropped for simplicity (see Eq.~[S53] from~\cite{Michaels2019c}), this is described by the following fourth-order DE:
\begin{equation}\label{strutsDE}
(1+\partial_{yy})^2\bar{W}-\delta\left(\frac{y^2}{2}\bar{W}'\right)'=0,
\end{equation}
for nondimensional horizontal co-ordinate and strain $\bar{x}$ and $\delta\ll 1$ on a filament anchored at $\bar{x}=\pm\bar{L}/2$. This is a Turing-Swift-Hohenberg-type equation, with the leftmost operator featuring frequently in equations describing pattern formation.

Since CGO RG was not practical (see below), Eq.~\eqref{strutsDE} was solved in~\cite{Michaels2019c} using MMS, requiring the extremely difficult guess of the existence of a hidden $\delta^{1/4}$(!) scale. However, it can be solved much more straightforwardly using hidden scale symmetry, as follows. Expanding $\bar{W}$ in $\delta$ as $\bar{W}=\bar{W}^{(0)}+\delta\bar{W}^{(1)}+O(\delta^2)$ yields perturbation equations:
\begin{equation}
(1+\partial_{yy})^2\bar{W}^{(0)}=0,\qquad (1+\partial_{yy})^2\bar{W}^{(1)} = \left(\frac{y^2}{2}\bar{W}^{(0)}{}'\right)'.
\end{equation}
The general solution with undetermined constants of integration for the zeroth-order term is:
\begin{equation}
\bar{W}^{(0)}(y)=A_1\cos y+A_2\sin y+\alpha_1 y\cos y+\alpha_2y\sin y.
\end{equation}
Painting divergent instances of $y$ in this reference solution and in its first-order derivative yields:
\begin{subequations}
	\begin{align}
	\bar{W}^{(0)}(y,\mu)&=A_1\cos y+A_2\sin y+\alpha_1 \mu\cos y+\alpha_2\mu\sin y\\
	\bar{W}^{(0)}{}'(y,\mu)&=-A_1\sin y+A_2\cos y+\alpha_1 (\cos y-\mu\sin y)+\alpha_2(\sin y+\mu\cos y).
	\end{align}
\end{subequations}
Thanks to demanding independence from $y$, no higher-order derivatives of the reference solution are needed to fully determine the symmetries of $\mu$ and the integration constants to $O(\delta)$. Applying the FT equations Eq.~\eqref{DearRGeq}-\eqref{DearRGeq2} to these yields the relations:
\begin{subequations}
	\begin{align}
	&A_1'+\alpha_1+\alpha_1'\mu= A_2'+\alpha_2+\alpha_2'\mu=O(\delta)\\
	&-A_1'-\alpha_1-\alpha_1'\mu+\alpha_2'= A_2'+\alpha_1'+\alpha_2+\alpha_2'\mu=O(\delta).
	\end{align}
\end{subequations}
From this we see that $\alpha_i'=O(\delta)$, and that therefore $A_i'=-\alpha_i+O(\delta)$ and consequently $A_i''=O(\delta)$. Since WLOG $A_i\neq c\mu+f(\delta,\mu)$, this implies $A_i'=-\alpha_i=O(\delta^q)$ where $q>0$. For $\bar{W}=O(1)$ it is therefore necessary that $A_i=O(1)$.

This interesting result demonstrates that all components depending on $\alpha_i$ in $\delta\bar{W}^{(1)}$ will in fact be $O(\delta^{1+q})$ and can be neglected in front of components depending on $A_i$. This allows solving the inhomogeneous ODE for $\bar{W}^{(1)}$ using only the terms $A_1\cos y+A_2\sin y$ for $\bar{W}^{(0)}$, greatly simplifying the problem:
\begin{equation}
(1+\partial_{yy})^2\bar{W}^{(1)}=-\frac{y^2}{2}(A_1\cos y+A_2\sin y)+y(-A_1\sin y+A_2\cos y).
\end{equation}

This is solved by:
\begin{align}
\bar{W}^{(1)}&=\frac{1}{192}\left[192(c_1+yc_2)+12 A_2 y + A_1(-9+6y^2+2y^4)\right]\cos y\nonumber\\
&\quad+\frac{1}{192}\left[192(c_3+yc_4)-12 A_1 y + A_2(-9+6y^2+2y^4)\right]\sin y.
\end{align}
There are any number of ways the integration constants $c_i$ can be chosen to eliminate divergences arising from $y$, rendering CGO RG impractical. Instead, painting all divergent terms in $\bar{W}^{(1)}$ and $\bar{W}^{(1)}{}'$, applying Eqs~\eqref{DearRGeq}-\eqref{DearRGeq2} to $\bar{W}^{(0)}+\delta\bar{W}^{(1)}$ and its derivative, and insisting on independence from $y$, leads immediately to the following four FT equations to $O(\delta^{1+q})$:
\begin{subequations}
	\begin{align}
	0&=A_1'+\alpha_1'\mu+\alpha_1+\frac{\delta}{48}\left[48c_2+3A_2+A_1\mu(3+2\mu^2)\right]\\
	0&=A_2'+\alpha_2'\mu+\alpha_2+\frac{\delta}{48}\left[48c_4-3A_1+A_2\mu(3+2\mu^2)\right]\\
	0&=\alpha_1'+A_2'+\alpha_2'\mu+\alpha_2+\frac{\delta}{48}\left[48c_4+6A_1\mu^2+A_2\mu(3+2\mu^2)\right]\\
	0&=\alpha_2'-(A_1'+\alpha_1'\mu+\alpha_1)+\frac{\delta}{48}\left[-48c_2+6A_2\mu^2-A_1\mu(2+2\mu^2)\right].
	\end{align}
\end{subequations}
Adding and subtracting these gives:
\begin{subequations}
	\begin{align}
	\alpha_1'&=-\frac{\delta}{48}\left(3+6\mu^2\right)A_1+O(\delta^{1+q})\\
	\alpha_2'&=-\frac{\delta}{48}\left(3+6\mu^2\right)A_2+O(\delta^{1+q}).
	\end{align}
\end{subequations}

\begin{figure}
	\centering
	\includegraphics[width=\textwidth]{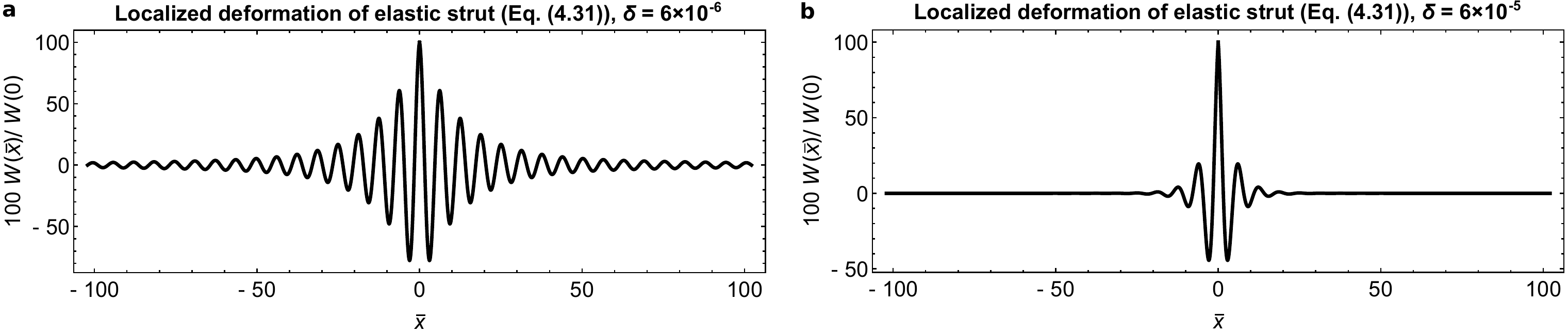}
	\caption{Plots of Eq.~\eqref{strutssoln} (with $\bar{x}=y+\bar{L}/2$) for two different values of strain, describing the growth-induced vertical buckling of an originally horizontal filament supported at each end and anchored in an elastic medium. \textbf{a}: At small values of excess strain, the filament buckles into an approximately sinusoidal shape, with only a modest amplitude modulation. \textbf{b} Larger values of excess strain are accommodated by much stronger amplitude modulation, resulting in strongly localized buckling around the centre of the filament.}
	\label{Fig6}
\end{figure}

In combination with the relations $A_i'=-\alpha_i+O(\delta)$, and keeping only the most divergent terms, this finally gives the amplitude equations:
\begin{subequations}
	\begin{align}
	A_1''&=\frac{\delta}{16}(2\mu^2+1)A_1+O(\delta^{1+q})\\
	A_2''&=\frac{\delta}{16}(2\mu^2+1)A_2+O(\delta^{1+q}).
	\end{align}
\end{subequations}
It also verifies $A_i'=O(\delta^{1/2})=-\alpha_i$, and thus that $q=1/2$. These equations are almost the same as the MMS equations in ref.~\cite{Michaels2019c} (and are identical if we consider only the most $\mu$-divergent terms). Integrating them from $y$ to 0 yields almost identical parabolic cylinder functions ($D[i,j]$) for $A_1$ and $A_2$. The symmetry of the system around $x=0$ provides the boundary conditions $\bar{W}'(\bar{x}=0)=\bar{W}'''(\bar{x}=0)=0$; imposing them and requiring the solution to be real yields:
\begin{equation}\label{strutssoln}
\bar{W}(\bar{x})=\tilde{A}_2\, D\!\!\left[-\frac{1}{2}+\frac{2^{1/2}}{16}\delta^{1/2},\frac{\delta^{1/4}}{2^{1/4}}(\bar{x}-\bar{L}/2)\right]\cos(\bar{x}).
\end{equation}
This is almost Eq.~(3.5) of ref.~\cite{Michaels2019c} but with a $\delta^{1/2}$ correction to the first argument of $D$. This correction arises from the greater simplicity of the hidden scale symmetry method, which permits us to keep the term proportional to $\bar{W}'$ in Eq.~\eqref{strutsDE}. In ref.~\cite{Michaels2019c}, by contrast, it was necessary to drop this term to perform the MMS calculations. We plot Eq.~\eqref{strutssoln} in Fig.~\ref{Fig6}; it describes the slow amplitude modulation of sinusiodal buckling in the elastic filament. As strain increases the amplitude modulation becomes steeper and the buckling becomes more localized towards the centre of the filament.

\subsection{CGO RG and RG/E are indirect methods for exploiting hidden scale symmetry}\label{sec:derivingCGO}

Eq.~\eqref{DearRGeq} resembles the CGO RG equation (Eq.~S1; see SI Sec.~S1 B for an introduction to CGO RG). It differs only in that the differentiation is with respect to $x$ rather than $x_0$ or $\mu$, and in that the limit $x\to x_0$ has been replaced by the limit $\epsilon\to 0$. This suggests that CGO RG may have a basis in approximate solution symmetries connecting the redundant parameter $x_0$ and $A_i$. This can be true only if $x_0$ is introduced in a manner such that the divergent series remains a valid perturbative expansion with arbitrary integration constants and yields an exact, non-divergent special solution when $x\to x_0$. Otherwise, its approximate symmetries are not symmetries of the exact solution, and \textit{cannot} give rise to a globally valid approximate solution. Since under these circumstances CGO RG has been shown to find the envelope of the family of perturbation series with different $x_0$~\cite{Kunihiro1995}, we call these ``envelope symmetries''.

To prove that this is indeed the case, we now derive the CGO RG equation from the finite transformation equations for envelope symmetries. The generator $\bm{X}_k(x_0,A_i)$ of the $k$\textsuperscript{th}-order envelope symmetry may be written:
\begin{equation}\label{BCsym_general}
\bm{X}_{k}(x_0,A_i)=\frac{\partial}{\partial x_0}+\sum_{i=1}^n\sum_{j=0}^k\epsilon^j\xi_{A_i}^{(j)}\frac{\partial}{\partial A_i},
\end{equation}
where W.L.O.G.\ the element of the tangent vector in the direction of $x_0$ has been set to 1. An invariant $J_k(x,x_0,A_i)$ of this approximate symmetry satisfies $\bm{X}_{k}(x_0,A_i)J_k(x,x_0,A_i)=0$:
\begin{equation}
0=\bm{X}_k(x_0,A_i)J_k=\frac{\partial J_k}{\partial x_0}+\sum_{i=1}^n\sum_{j=0}^k\epsilon^j\xi_{A_i}^{(j)}\frac{\partial J_k}{\partial A_i}.
\end{equation}
From the method of characteristics this partial differential equation is reduced to $dJ_k/dx_0=0$, i.e.\ $J_k=$ const., along characteristics defined by:
\begin{equation}
\sum_{j=0}^k\epsilon^j\xi_{A_i}^{(j)}=\frac{dA_i}{dx_0}.
\end{equation}
Now, using this result in Eq.~\eqref{BCsym_general} yields:
\begin{equation}
\bm{X}_k(x_0,A_i)=\frac{\partial }{\partial x_0}+\sum_i\frac{dA_i}{dx_0}\frac{\partial}{\partial A_i}.
\end{equation}
If we have not already computed the envelope symmetry then $\xi_{A_i}^{(j)}$ and thus $dA_i/dx_0$ are not known. They may, however, be in principle calculated by observing that the perturbation series to order $k,\ y_k$, is left invariant by the envelope symmetry only in the limit $x\to x_0$, i.e. $\bm{X}_k(x_0,A_i)y_k|_{t\to t_0}=0$. Thus:
\begin{equation}\label{CGOFT}
\frac{\partial y_k}{\partial x_0}\bigg|_{x\to x_0}+\sum_i\frac{dA_i}{dx_0}\frac{\partial y_k}{\partial A_i}\bigg|_{x\to x_0}=0,
\end{equation}
which is none other than the CGO RG equation. 

This provides a highly intuitive explanation for \textit{why} the seemingly trivial CGO RG equation should give globally valid approximate solutions: it is equivalent to the finite transformation (FT) equations, valid globally, for the approximate symmetry connecting integration constants $A_i$ and initial value $x_0$. When integrated from a known point on the exact solution manifold, the resultant finite transformation or orbit of this symmetry is necessarily a global approximate solution manifold of the desired order in $\epsilon$. This also implies that the ``envelopes'' that form the mathematical heart of RG/E~\cite{Kunihiro1995,Kunihiro1997,Kunihiro_book} are exactly equivalent to the orbits of these approximate symmetries. In support of this viewpoint, it has previously been observed (although not to our knowledge proved generally) that the CGO RG equation has the structure of an asymptotic expansion of a Lie group generator~\cite{Goto1999}. The meaning of this result in terms of approximate symmetry properties of the exact solution was not investigated at the time. 

What is the precise relationship between hidden scale symmetries and the envelope symmetries that we have shown underlie CGO RG and its alternative formulation, RG/E? To answer this, we will use again the overdamped linear oscillator, transformed into the inner layer (Eq.~\eqref{overdampedHO}):
\begin{equation}
\frac{d^2y}{d\tau^2}+\frac{dy}{d\tau}+\epsilon y=0.
\end{equation}
and its first order perturbative solution Eq.~\eqref{Eq33}:
\begin{equation}
y_1(\tau)=A+Be^{-\tau}+\epsilon\left[-A\tau +B\tau e^{-\tau}\right],
\end{equation}
Now, in the original CGO formulation, one is free to ``split'' the divergent terms using $\tau_0$ however one sees fit, and redefine $A,\ B$ to absorb the resulting terms dependent on $\tau_0$ only, to ensure the $\tau$-dependent terms vanish at $\tau=\tau_0$. $A\tau$ can only become $\tau-\tau_0$. For $B\tau e^{-\tau}$, however, there are an infinite number of choices, and in its most general form the split perturbation series becomes:
\begin{equation}\label{splitted}
y_1(\tau,\tau_0)=A+Be^{-\tau}+\epsilon\left[-A(\tau-\tau_0) +B(\tau e^{-\tau}-s\tau_0 e^{-\tau_0}-(1-s)\tau_0 e^{-\tau})\right],
\end{equation}
where $s$ can take any value.

Applying the CGO RG equation (Eq.~\eqref{CGOFT}), for arbitrary $s$, yields:
\begin{multline}
\left(A'+B'e^{-\tau}+\epsilon\left[A-A'(\tau-\tau_0)+B(-s e^{-\tau_0}+s\tau_0 e^{-\tau_0}-(1-s) e^{-\tau})\right.\right.\\\left.\left.+B'(\tau e^{-\tau}-s\tau_0 e^{-\tau_0}-(1-s)\tau_0 e^{-\tau})\right]\right)\bigg|_{\tau\to\tau_0}=0.
\end{multline}
Now the limit $\tau\to\tau_0$ is invariably used in the literature to remove divergent terms proportional to $\tau-\tau_0$. In CGO RG, to prevent underdetermination, independence of the integration constants from $\tau$ is also demanded. However, since the integration constants are assumed to have $\tau_0$ dependence and since in this limit $\tau$ is no longer independent from $\tau_0$, this is apparently inconsistent. Instead we are left with the doubly underdetermined equation:
\begin{equation}\label{underdetermined}
A'+B'e^{-\tau_0}+\epsilon\left[A+B(s\tau_0 e^{-\tau_0}- e^{-\tau_0})\right]=0,
\end{equation}
where $A(\tau_0), B(\tau_0)$ and $s$ are unknown. The correct solution (Eq.~\eqref{HSsymsoln}; Fig.~\ref{Fig3}\textbf{a}) can be arrived at by setting $s=0$ and then partitioning terms in $A$ and terms in $B$ into separate equations~\cite{Chen1996}. But how can we justify these steps mathematically?

To make progress we first remark that Eq.~\eqref{underdetermined} generates the orbits of not one but a family of envelope symmetries satisfied by the perturbation series, only one member of which can be a symmetry of the exact solution. Each splitting $s$ can thus be interpreted as a different way to parametrize the perturbation series into a family of adjacent curves, each with its own infinite set of envelopes, only one of which provides a valid global solution. Since the final step is to replace $\tau_0$ with $\tau$, this must be the one corresponding to the hidden scale symmetry. As in the hidden scale method, to identify it we must consider also the derivative of the splitted perturbation series Eq.~\eqref{splitted}:
\begin{equation}\label{splittedderiv}
\frac{dy_1(\tau,\tau_0)}{d\tau}=-Be^{-\tau}+\epsilon\left[-A +B( e^{-\tau}-\tau e^{-\tau}+(1-s)\tau_0 e^{-\tau})\right].
\end{equation} 
The reason that only $s=0$ gives the correct splitting is now revealed to be that only when $s=0$ does the divergence in both Eq.~\eqref{splitted} and Eq.~\eqref{splittedderiv} vanish as $\tau\to\tau_0$. Setting $s=0$ and applying the CGO RG equation (Eq.~\eqref{CGOFT}) then yields $A'=B'=O(\epsilon)$ and thus, in combination with Eq.~\eqref{underdetermined}:
\begin{equation}\label{splittedderivRG}
	A'+B'e^{-\tau_0}+\epsilon\left[A-Be^{-\tau_0}\right]=0,\qquad -B'e^{-\tau_0}+\epsilon\left[Be^{-\tau_0}\right]=0.
\end{equation}
By eliminating underdetermination, this gives the desired separation of terms in $A$ and $B$, recovering the equations in ref.~\cite{Chen1996}.

Thus, CGO RG and RG/E are effectively indirect methods to identify and exploit hidden scale symmetries, requiring introduction of an arbitrary $\tau_0$ or $x_0$. CGO RG yields the envelope symmetries of the perturbation series; the one that is an inverted hidden scale symmetry and therefore yields a correct solution must be identified by trial-and-error and by guesswork. By contrast, in RG/E one is meant to convert $n$\textsuperscript{th}-order ODEs into $n$ 1\textsuperscript{st}-order ODEs prior to application of the RG equation. This prevents underdetermination analogously to imposing Eq.~\eqref{DearRGeq2} and also avoids potential inconsistencies introduced by requiring independence from $\tau$, leading to an inverted hidden scale symmetry without guesswork.

\subsection{Key ``singular'' problems are actually regular boundary condition perturbations}\label{sec:switchback}

An important apparent success of the CGO RG method is its generation of unprecedentedly accurate solutions to the challenging ``switchback'' family of problems, with vastly less effort than was required using older approaches. However, the derivations of these solutions appear inconsistent with a Lie group origin for CGO RG, and incompatible with RG/E~\cite{Kunihiro1995,Kunihiro1997}. Here, we resolve this apparent paradox by showing that these problems are in fact solved by \textit{regular} perturbation of the boundary conditions. 

The method of matched asymptotic expansions, a widely used singular perturbation technique, was first given mathematical foundations in the 1950s by Kaplun and Lagerstrom~\cite{Faria2017}, and further developed by them and others in the 1960s~\cite{Lagerstrom1972}, with the help of a family of particularly challenging model problems, known as switchback problems, that are asymptotically similar to low Reynolds number problems. A minimal model for them is given by~\cite{Veysey2007}:
\begin{subequations}\label{badworseterrible}
	\begin{equation}
	\frac{d^2u}{dx^2}+\frac{n-1}{x}\frac{du}{dx}+u\frac{du}{dx}+\delta\left(\frac{du}{dx}\right)^2=0,
	\end{equation}
	\begin{equation}
	u(x=\epsilon)=0,\qquad u(x=\infty)=1.
	\end{equation}
\end{subequations}
$n$ is interpretable as the dimension of the space, $\epsilon$ the radius of the sphere or circle inserted at the origin, and $u(x=\infty)=1$ the velocity of the undisturbed fluid flow. These problems with $n=2,3$ and $\delta=0,1$, and the outer variable $r=x/\epsilon$, were also employed in a pedagogical context as a ``worst-case scenario'' for the difficulty of applying matched asymptotic expansions in Hinch's textbook~\cite{Hinch1991}.

In ref.~\cite{Chen1994}, Eq.~\eqref{badworseterrible} in inner variable $x$ with $n=3,\ \delta=0$ was solved using CGO RG, with the intent to demonstrate the advantages of CGO RG over matched asymptotics. In ref.~\cite{Chen1996}, the same was done for Eq.~\eqref{badworseterrible} with $n=2,\ \delta=0,1$; in ref.~\cite{Veysey2007} this calculation was revisited in more detail. Indeed, the solutions found were both more accurate than the matched asymptotic solution of Hinch~\cite{Hinch1991}, and calculated far more easily.

In all cases, the reference system could not be solved fully, since setting $\epsilon=0$ simplifies neither the inner layer DE nor the boundary conditions. (The sphere or circle does not actually vanish, as the $u=0$ boundary condition remains even in this limit; instead it becomes a no-slip point disturbance.) Instead, the partial reference solution $u^{(0)}=A=$ const.\ was used as a starting point to construct a \textit{partial} perturbation series in $\epsilon$, by expanding $u(x,\epsilon)=A+\sum_{i=1}\epsilon^i u^{(i)}(x)$, yielding at first order $u(x)=u_1(x,\epsilon)+O(\epsilon^2)$, where:
\begin{subequations}\label{fakepertseries}
	\begin{align}
	u_1(x)= A+\epsilon Be_2(x) & \qquad (n=3,\ \delta=0) \\
	u_1(x)= A+\epsilon Be_1(x) & \qquad (n=2,\ \delta=0,1)
	\end{align}
\end{subequations}
where $e_n(t)=\int_{t}^{\infty}d\rho \rho^{-n}e^{-\rho}$.
These are not valid series solutions as they do not satisfy the boundary conditions at each order. Nonetheless, CGO RG was successfully applied, yielding:
\begin{subequations}\label{badterriblesolns}
	\begin{align}
	u(x)= 1-\frac{e_2(x)}{e_2(\epsilon)}+O(\epsilon^2) &\qquad (n=3,\ \delta=0) \label{badsoln}\\
	u(x)= 1-\frac{e_1(x)}{e_1(\epsilon)}+O(\epsilon^2) &\qquad (n=2,\ \delta=0,1)\label{terriblesoln}
	\end{align}
\end{subequations}
Note these are the same as the original perturbation series with constants $A,B$ chosen to match the boundary conditions. Since these choices convert the $O(\epsilon)$ terms to $O(1)$ terms, Eqs.~\eqref{fakepertseries} are not even valid perturbation series in $\epsilon$. These successes would thus appear to violate our Lie theoretic principle that CGO RG can only be applied to valid perturbation series solutions.

\begin{figure}
	\centering
	\includegraphics[width=0.92\textwidth]{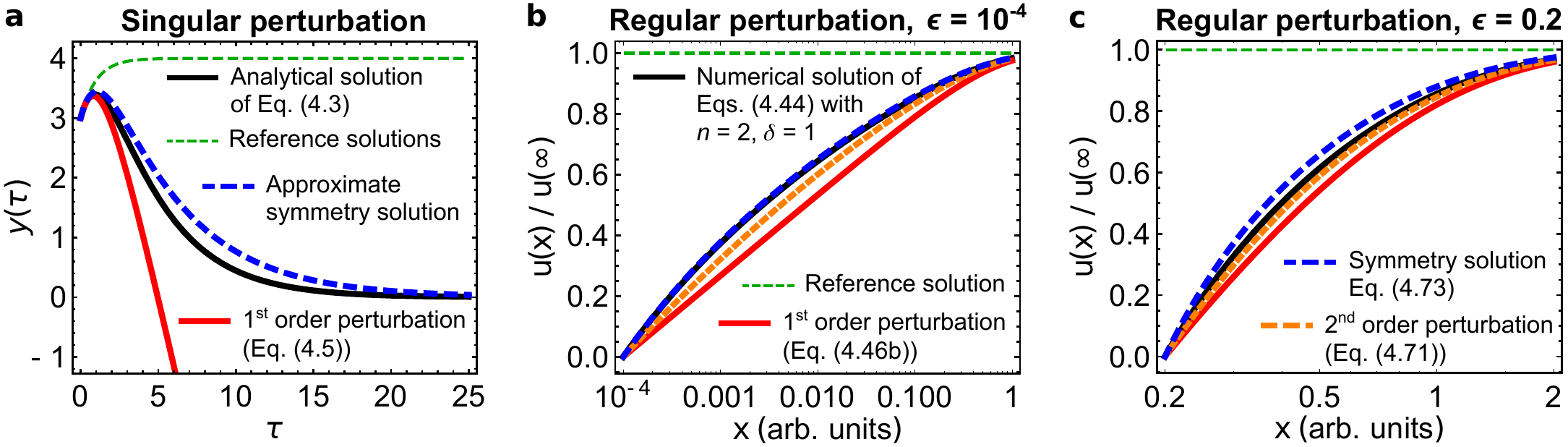}
	\caption{Regular vs singular perturbation problems. \textbf{a}: The overdamped harmonic oscillator yields a singular perturbation series (Eq.~\eqref{Eq33}). This diverges for values of $\tau$ sufficiently far from where initial conditions were imposed. It is easier to regularize this using the hidden scale FT equation Eq.~\eqref{DearRGeq} than the CGO RG equation. Shown: $\epsilon=0.2,\ y(0)=3,\ y'(0)=1$. \textbf{b}-\textbf{c}: By contrast, the ``switchback'' equations, Eqs.~\eqref{badworseterrible}, yield regular perturbation series in a switching parameter $a$ for the inner boundary condition (Eqs~\eqref{badworseterrible_pert}-\eqref{badterriblesolns1st}). These series converge globally on the exact solution as higher-order terms are added, even when $a=1$. Shown: $n=2,\ \delta=1,\ a=1$. \textbf{b}: The hidden scale FT equation yields the exact $x\to 0$ asymptotic solution Eq.~\eqref{terriblesoln2o} (shown: $\epsilon=10^{-4}$). At very small $\epsilon$, this is the most accurate approximation of Eqs~\eqref{terrible_pert}. \textbf{c}: At larger but still small $\epsilon$, the second order perturbative solution is instead the more accurate.}
	\label{Fig3}
\end{figure}

Trusting in our Lie theoretic interpretation of CGO RG implies that these partial series must still be valid perturbation series solutions despite appearances, just indexed by a different perturbation parameter $a$. For this to be true $u_0=A$ (unimpeded flow) must satisfy the boundary conditions at zeroth order. This in turn requires that the perturbation parameter $a$ enters the boundary conditions rather than the DEs, in a way that $a=0$ reduces both boundary conditions to $A=1$. So, the switchback problem can be formulated as a ``boundary condition perturbation'':
\begin{subequations}\label{badworseterrible_pert}
	\begin{equation}
	\frac{d^2u}{dx^2}+\frac{n-1}{x}\frac{du}{dx}+u\frac{du}{dx}+\delta\left(\frac{du}{dx}\right)^2=0,
	\end{equation}
	\begin{equation}
	u(x=\epsilon)=1-a,\qquad u(x=\infty)=1.
	\end{equation}
\end{subequations}
Expanding $u=u^{(0)}+au^{(1)}+O(a^2)$ then yields the zeroth order perturbation equation:
\begin{subequations}\label{badworseterrible_zerothDE}
	\begin{equation}
	\frac{d^2u^{(0)}}{dx^2}+\frac{n-1}{x}\frac{du^{(0)}}{dx}+u^{(0)}\frac{du^{(0)}}{dx}+\delta\left(\frac{du^{(0)}}{dx}\right)^2=0,
	\end{equation}
	\begin{equation}
	u^{(0)}(x=\epsilon)=1,\qquad u^{(0)}(x=\infty)=1,
	\end{equation}
\end{subequations}
which is solved by $u^{(0)}=1$. The first order equation is then:
\begin{subequations}\label{badworseterrible_firstDE}
	\begin{equation}
	\frac{d^2u^{(1)}}{dx^2}+\frac{n-1}{x}\frac{du^{(1)}}{dx}+u^{(0)}\frac{du^{(1)}}{dx}=0,
	\end{equation}
	\begin{equation}
	u^{(1)}(x=\epsilon)=-1,\qquad u^{(1)}(x=\infty)=0,
	\end{equation}
\end{subequations}
which is solved by:
\begin{subequations}\label{badterriblesolns1st}
	\begin{align}
	u^{(1)}(x)= -\frac{e_2(x)}{e_2(\epsilon)}+O(\epsilon^2) &\qquad (n=3,\ \delta=0) \label{badsoln1st}\\
	u^{(1)}(x)= -\frac{e_1(x)}{e_1(\epsilon)}+O(\epsilon^2) &\qquad (n=2,\ \delta=0,1)\label{terriblesoln1st}
	\end{align}
\end{subequations}
The solutions Eqs.~\eqref{badterriblesolns} are then revealed as the first order perturbation series in $a$, $u^{(0)}+au^{(1)}$, with $a=1$ to recover the desired boundary condition. Furthermore, plotting the numerical solution to the ``terrible'' problem (Eq.~\eqref{badworseterrible} with $n=2,\ \delta=1$) alongside its first- and second-order perturbative solutions in $a$ (Eqs.~\eqref{terriblesoln} and \eqref{2pert}) with $a=1$ unambiguously confirms the regular nature of this problem, and that the radius of convergence of the perturbation series is $>1$ (Fig.~\ref{Fig3}\textbf{b}-\textbf{c}). That this is a \textit{regular} perturbation in the boundary condition is also proved explicitly in SI Sec.~S3.

\subsection{Regular boundary condition perturbation of the Navier-Stokes equations}\label{sec:Oseen}
We suspect that the identification of \textit{regular} perturbations in the boundary conditions as singular problems may have occurred frequently in the past (e.g.\ the problem studied in ref.~\cite{Clark2023}); their re-evaluation may resolve several apparent paradoxes in the literature. An example of significance is the incompressible steady Navier-Stokes equation of fluid mechanics:
\begin{equation}\label{NavierStokes}
u\cdot\nabla u = \nu \nabla^2u-\frac{1}{\rho}\nabla p.
\end{equation}
$u$ and $p$ are the velocity and pressure fields, $\rho$ the liquid density, and $\nu\propto 1/R$ the kinematic viscosity, for Reynolds number $R$. The Oseen equation is a famous and highly successful low-Reynolds number approximation of this equation:
\begin{equation}
U\cdot\nabla u = \nu \nabla^2u-\frac{1}{\rho}\nabla p,
\end{equation}
with boundary conditions ($U$ is the (constant) bulk fluid velocity):
\begin{equation}
	u(r)=0\quad r \in \text{ surface of fixed body},\quad\lim_{|r|\to\infty}u(r)=U,\quad\lim_{|r|\to\infty}p(r)=p_\infty.
\end{equation}
The Oseen equation was originally obtained from Eq.~\eqref{NavierStokes} by replacing $u$ with $U+u^{(1)}$, without, in our view, a clear view of a perturbation parameter. Our work in the preceding section, however, implies that the correct perturbation parameter to use is $a$, where the first boundary condition has been modified to:
\begin{equation}
u(r)=U-a\quad r \in \text{{surface of fixed body}}.
\end{equation}
Now the Oseen equation is a true first order perturbation equation, and the perturbation series can match the boundary conditions precisely at each order in $a$.

It has been observed~\cite{Roper2009} that only for $R<\sim 1$ is this approximation valid near the body surface. This is presumably because the radius of convergence of the resultant perturbation series is dependent on $R$ (and likely also on the body geometry). Only for $R<\sim 1$ does the radius of convergence exceed $U$, so that the imposition of the no-slip boundary condition $u=0$ at the body surface qualifies as a regular perturbation. This would also explain why the Oseen equation is always a valid approximation far enough from the body. For any $R$ there should exist a surface $r^*$ in the flow sufficiently far from the body that the exact $u(r^*)$ can be formulated as a perturbed ``surface'' boundary condition $u(r^*)=U-a$ where $a$ is within the radius of convergence of the regular perturbation series for the particular choice of $R$.

\subsubsection{Navier-Stokes equation for the infinite cylinder as a boundary condition perturbation}
To illustrate in more detail how the Oseen equation can give rise to regular perturbative solutions to the Navier-Stokes equations, we consider the classic problem of flow past an infinite cylinder, with ``no-slip'' boundary conditions at the cylinder surface (remarking that flow past a sphere can be treated in a similar fashion). Cylindrical coordinates are logical in this instance, and the two-dimensional nature of the problem then permits its formulation in terms of the Lagrange stream function $\psi$~\cite{Veysey2007}, defined as:
\begin{equation}
u_r=\frac{1}{r}\frac{d\psi}{d\theta},\quad u_\theta=-\frac{d\psi}{dr},\quad u_z=0.
\end{equation}
Choosing the scaling $\Psi=R\psi,\ \rho=Rr$, the steady incompressible Navier-Stokes equation then becomes~\cite{Veysey2007}:
\begin{equation}\label{full}
\nabla_\rho^2\nabla_\rho^2\Psi=\frac{1}{\rho}\left(\frac{\partial\Psi}{\partial\theta}\frac{\partial}{\partial\rho}-\frac{\partial\Psi}{\partial\rho}\frac{\partial}{\partial\theta}\right)\nabla_\rho^2\Psi.
\end{equation}
The desired ``no-slip'' BCs are:
\begin{equation}
	\lim\limits_{\rho\to\infty}\Psi=\rho\sin\theta,\qquad \Psi(\rho=R,\theta)=0,\qquad 	u_\theta=-\frac{\partial\Psi}{\partial\rho}\bigg|_{\rho=R}=0.
\end{equation}
Note:
\begin{equation}
\nabla_\rho^2=\frac{\partial^2}{\partial\rho^2}+\frac{1}{\rho}\frac{\partial}{\partial\rho}+\frac{1}{\rho^2}\frac{\partial^2}{\partial\theta^2}.
\end{equation}
These BCs can be rewritten in terms of perturbation parameter $a$, to be later set to 1:
\begin{equation}
	\lim\limits_{\rho\to\infty}\Psi=\rho\sin\theta,\qquad \frac{\partial\Psi}{\partial\rho}\bigg|_{\rho=R}=(1-a)\sin\theta,\qquad \Psi(\rho=R,\theta)=(1-a)R\sin\theta.
\end{equation}
Note, other boundary conditions can also be used so long as they recover the outer BC at $a=0$. For instance, ``slip'' boundary conditions could be set up as above, but with $(1-as)\sin\theta$ for the second BC, and $s<1$, such that at $a=1$ no flow into or out of the solid body is allowed, but some slippage around its edge occurs.

\subsubsection{Zeroth order perturbative solution}
Expanding in $a$, i.e.\ $\Psi=\Psi^{(0)}+a\Psi^{(1)}+a^2\Psi^{(2)}+\dots$, at zeroth order we have Eq.~\eqref{full} but for $\Psi^{(0)}$, solved with BCs:
\begin{equation}
	\lim\limits_{\rho\to\infty}\Psi^{(0)}=\rho\sin\theta,\qquad \frac{\partial\Psi^{(0)}}{\partial\rho}\bigg|_{\rho=R}=\sin\theta,\qquad \Psi^{(0)}(\rho=R,\theta)=R\sin\theta.
\end{equation}
The only solution to this equation that can satisfy both inner and outer BCs is:
\begin{equation}\label{zerothorderOseen}
\Psi^{(0)}=\rho\sin\theta.\qquad(\text{Note, }\nabla_\rho^2\Psi^{(0)}=0.)
\end{equation}
If the Navier-Stokes equation is treated as a singular perturbation problem in the Reynolds number, this would only be a partial zeroth-order solution~\cite{Veysey2007}. However,  Eq.~\eqref{zerothorderOseen} is the full zeroth-order solution for any surface boundary condition perturbation that recovers the outer BC at $a=0$, circumventing this conceptual difficulty.

\subsubsection{(Regular) first order equation is the Oseen equation}

Using this $\Psi^{(0)}$, the first-order perturbation equation simplifies to:
\begin{equation}\label{Oseen2}
\nabla_\rho^2\nabla_\rho^2\Psi^{(1)}-\frac{1}{\rho}\left(\rho\cos\theta\frac{\partial}{\partial\rho}-\sin\theta\frac{\partial}{\partial\theta}\right)\nabla_\rho^2\Psi^{(1)}=0,
\end{equation}
which is identical to the Oseen equation. But the correct boundary conditions are:
\begin{equation}\label{BCOseen2}
	\lim\limits_{\rho\to\infty}\frac{\Psi^{(1)}}{\rho}=0,\qquad \frac{\partial\Psi^{(1)}}{\partial\rho}\bigg|_{\rho=R}=-\sin\theta,\qquad\Psi^{(1)}(\rho=R,\theta)=-R\sin\theta.
\end{equation}
Thus, crucially, the Oseen equation is the \textit{regular} first-order boundary condition perturbation expansion for the Navier-Stokes equation, satisfied by and solved for $\Psi^{(0)}+a\Psi^{(1)}$ with $a=1$.

Solving Eq.~\eqref{Oseen2} and imposing the boundary conditions at $\rho\to\infty$ gives~\cite{Veysey2007,Tomotika1950}:
\begin{align}\label{Oseensemigensoln}
\Psi^{(1)}&=\sum_{n=1}^{\infty}\left[B_n\rho^{-n}+\sum_{m=0}^{\infty}X_m\rho\,\Phi_{m,n}\!\left(\frac{\rho}{2}\right)\right]\sin n\theta,\\
\Phi_{m,n}&=(K_{m+1}+K_{m-1})(I_{m+n}+I_{m-n})+K_m(I_{m+n+1}+I_{m-n+1}+I_{m+n-1}+I_{m-n-1}),
\end{align}
where $I_m(z)$ and $K_m(z)$ are modified Bessel functions of the first and second kind, respectively. Eq.~(136) from ref.~\cite{Veysey2007} is thus $\Psi^{(0)}+a\Psi^{(1)}$ with $a=1$, i.e.\ an exact regular first order perturbation in the boundary conditions of the Navier-Stokes equation. For small enough $R$ we can set $a=1$ and remain within the radius of convergence, as discussed, explaining the success of the Oseen equation. Imposing the $\rho=R$ BC, Eq.~\eqref{Oseensemigensoln} reduces further to (see SI Sec.~S4):
\begin{align}
\Psi^{(1)}&=-\frac{R^2}{\rho}\sin\theta +\sum_{n=1}\sum_{m=0}X_m \left[\rho\,\Phi_{m,n}\!\left(\frac{\rho}{2}\right)-\frac{R^{n+1}}{\rho^n}\,\Phi_{m,n}\!\left(\frac{R}{2}\right)\right]\sin n\theta,\label{BCsoln1}\\
A_{nm}X_m&=-2\delta_{n,1},\ n>0,\ m\geq 0,\quad A_{nm}=(n+1)\Phi_{m,n}\!\left(\frac{R}{2}\right)+\frac{R}{2}\,\Phi_{m,n}'\!\left(\frac{R}{2}\right).\label{mateq}
\end{align}

\subsubsection{Analytical $R\to 0$ solution}
Although exact, Eq.~\eqref{BCsoln1}, like Eq.~\eqref{Oseensemigensoln}, is unusable. This is because although
all integration constants are determined by the boundary conditions, this requires inverting an infinite dimensional matrix (e.g.\ Eq.~\eqref{mateq}). In practice a finite truncation of the infinite terms of Eq.~\eqref{BCsoln1} must be made~\cite{Veysey2007,Tomotika1950,Gustafsson2013}.

As discussed in ref.~\cite{Veysey2007}, the only consistent truncation in the $R\to 0$ limit is to retain only the first harmonic in Eq.~\eqref{BCsoln1}, setting $X_m=0\ \forall m>0$, and solving Eq.~\eqref{mateq} truncated to the first row and column. This yields:
\begin{align}
X_0=&\frac{-4}{4\Phi_{0,1}\!\left(\frac{R}{2}\right)+R\,\Phi_{0,1}'\!\left(\frac{R}{2}\right)},\quad X_{m>0}=0.
\end{align}
Veysey's solution (Eq.~(146) of ref.~\cite{Veysey2007}) follows from Eq.~\eqref{BCsoln1}, albeit with a factor of $R$ difference due to their expansion being putatively in $R$ rather than $a$. Consequently this represents the $R\to 0$ asymptotic solution to the Oseen equation, and thus the $R\to 0$ asymptotic limit of the regular perturbation solution to the Navier-Stokes equations.

Similarly to the switchback problems, the return of Eq.~(146) unmodified upon application of CGO RG in ref.~\cite{Veysey2007} reflects its status as a regular first order perturbative solution. Again, since Eq.~(146) is a truncation of a valid (boundary condition) perturbative solution, this apparent successful implementation of CGO RG does not contradict its Lie symmetry origin.

\subsection{Singular perturbation methods can find exact solutions if the underlying hidden scale symmetries are exact}\label{sec:exactsym}

The ``terrible'' problem (Eq.~\eqref{badworseterrible_pert} with $n=2,\ \delta=1$) is:
\begin{subequations}\label{terrible_pert}
	\begin{equation}
	\frac{d^2u}{dx^2}+\frac{1}{x}\frac{du}{dx}+u\frac{du}{dx}+\left(\frac{du}{dx}\right)^2=0,
	\end{equation}
	\begin{equation}
	u(x=\epsilon)=1-a,\qquad u(x=\infty)=1.
	\end{equation}
\end{subequations}
Expanding $u$ in $a$ as before, but adding a second order term, i.e.\ $u=u^{(0)}+au^{(1)}+a^2u^{(2)}+O(a^3)$, the second order perturbation equation is:
\begin{subequations}\label{badworseterrible_secondDE}
	\begin{equation}
	\frac{d^2u^{(2)}}{dx^2}+\frac{1}{x}\frac{du^{(2)}}{dx}+u^{(0)}\frac{du^{(2)}}{dx}=-u^{(1)}\frac{du^{(1)}}{dx}-\left(\frac{du^{(1)}}{dx}\right)^2,
	\end{equation}
	\begin{equation}
	u^{(2)}(x=\epsilon)=0,\qquad u^{(2)}(x=\infty)=0.
	\end{equation}
\end{subequations}
Solving Eq~\eqref{badworseterrible_zerothDE} as before to yield $u^{(0)}=1$, and solving Eq.~\eqref{badworseterrible_firstDE} for arbitrary boundary conditions, allows us to solve Eq.~\eqref{badworseterrible_secondDE}, yielding the regular second order perturbation series:
\begin{equation}\label{2pert}
u=1+a[A+B e_1(x) ]-\frac{a^2}{2}B^2e_1(x)^2+\mathcal{R}+O(a^3),
\end{equation}
where $\mathcal{R}$ are terms that diverge less rapidly as $x\to 0$ than $e_1(x)^2$ does. We chose to obtain arbitrary integration constants $A$ and $B$ at $O(a)$ not at $O(1)$ because $u$ everywhere is within $O(a)$ of 1. Using the original formulation of CGO RG there are at least three possible splittings of the most divergent second-order term:
\begin{subequations}
	\begin{align}
	\frac{1}{2}B^2e_1(x)^2&\to \frac{1}{2}B^2(e_1(x)^2-e_1(x_0)^2),\label{splite1}\\
	\frac{1}{2}B^2e_1(x)^2&\to \frac{1}{2}B^2(e_1(x)-e_1(x_0))^2,\label{splite2}\\
	\frac{1}{2}B^2e_1(x)^2&\to \frac{1}{2}B^2e_1(x)(e_1(x)-e_1(x_0)).
	\end{align}
\end{subequations}
Veysey \textit{et al} investigated the splittings Eqs.~\eqref{splite1}-\eqref{splite2} in ref.~\cite{Veysey2007}. It was found that CGO RG with splitting Eq.~\eqref{splite1} returns $A$ and $B$ as constants which are then determined by the boundary conditions. In other words, the perturbation series with $a=1$ is recovered unchanged, similarly to the first order case. Veysey \textit{et al} also found that using splitting Eq.~\eqref{splite2} yields instead a different expression:
\begin{equation}\label{terriblesoln2o}
u=\ln\left(e+(1-e)\frac{e_1(x)}{e_1(\epsilon)}\right).
\end{equation}
Note these calculations led to the correct results despite $\epsilon$ having been used incorrectly instead of $a$, because this constant error can be absorbed into $B$.

It was not explored in detail by Veysey \textit{et al} which expression offers the superior approximate solution, although it was stated that the splitting Eq.~\eqref{splite1} is the best approach. In SI Sec.~S3 we show that Eq.~\eqref{terriblesoln2o} is the exact sum of the most divergent terms in the perturbation series, to infinite order. So, with splitting Eq.~\eqref{splite2}, CGO RG has not identified an approximate solution, but instead the asymptotic $x\to \epsilon$ limit of the \textit{exact} solution to the ``terrible'' problem. Clearly, Eq.~\eqref{splite2} is thus the correct splitting, and Eq.~\eqref{terriblesoln2o} the best solution, but only when $\epsilon$ is small enough to justify dropping the least divergent terms at second order (Fig.~\ref{Fig3}\textbf{b}-\textbf{c}). 

Using instead our hidden scale symmetry method avoids this trial-and-error. Changing variables for simplicity to $\tau=e_1(x)$ and dropping the less-divergent terms, the perturbation series and its derivative become:
\begin{subequations}
	\begin{align}
	u(\tau)&=1+a[A+B\tau]-\frac{a^2}{2}B^2\tau^2+O(a^3),\\
	\dot{u}(\tau)&=aB-a^2B^2\tau+O(a^3).
	\end{align}
\end{subequations}
Applying the hidden scale finite transformation (FT) equation to these yields:
\begin{subequations}
	\begin{align}
	0&=A'+B+B'\tau-aB^2\tau-aBB'\tau^2+O(a^2),\\
	0&=B'-aB^2-2aBB'\tau+O(a^2).\label{DearRGterribledot}
	\end{align}
\end{subequations}
Since Eq.~\eqref{DearRGterribledot} gives $B'=O(a)$, these yield the finite transformation equations:
\begin{equation}
	0=B'-aB^2+O(a^2),\qquad 0=A'+B+O(a^2).
\end{equation}
Integrating from $(\tau,A,B)=(\tau,\tilde{A},\tilde{B})$ to $(0,A(\tau),B(\tau))$, yields:
\begin{subequations}\label{intRGAB}
	\begin{align}
	B(\tau)&=\frac{\tilde{B}}{1+a\tilde{B}\tau}+O(a^2),\\
	a(A(\tau)-\tilde{A})=&-a\int_{\tau}^{0}B(\tau)d\tau=\ln\left[1+a\tilde{B}\tau\right]+O(a^3).\label{Asoln}
	\end{align}
\end{subequations}
The convergent special solution is $u(\tau\to 0)=1+aA$, so the global solution is:
\begin{equation}
u(\tau)=1+a\tilde{A}+\ln\left[1+a\tilde{B}\tau\right].
\end{equation}
Switching variables back from $\tau$ to $x$, we have that $\tilde{A}=0$ from the boundary condition $u(x=\infty)=1$. $\tilde{B}$ is obtained by satisfying the boundary condition $u(x=\epsilon)=1-a$:
\begin{equation}
1-a=1+\ln\left[1+a\tilde{B} e_1(\epsilon)\right]\ \Rightarrow a\tilde{B} =\frac{e^{-a}-1}{e_1(\epsilon)}.
\end{equation}
Substituting into Eq.~\eqref{Asoln}, we recover the symmetry-transformed special solution $u(x)=1+aA(x)$:
\begin{equation}
u(x)=1+\ln\!\left[1-(1-e^{-a})\frac{e_1(x)}{e_1(\epsilon)}\right]+O(a^3).
\end{equation}
Setting $a=1$ yields finally the exact asymptotic solution Eq.~\eqref{terriblesoln2o} (Fig.~\ref{Fig3}\textbf{b}-\textbf{c}).

This is not the only example of CGO RG finding exact solutions (see also e.g.\ the nonlinear problem of Carrier, Eq.~(3.34) of ref.~\cite{Chen1996}). This can only happen if the underlying hidden scale symmetry is exact. To be identifiable order-by-order by CGO RG or our hidden scale method, it must also a finite power series in the perturbation parameter. Indeed, CGO RG to first order solves the problem of Carrier exactly in ref.~\cite{Chen1996} because the exact envelope symmetry is linear in $\epsilon$. In the present case, CGO RG or hidden scale symmetry analysis finds the asymptotic behaviour of the exact solution because the hidden scale symmetry is \textit{asymptotically} a first-order polynomial in $a$ as $x\to 0$, that can be discovered by CGO RG from the second-order perturbation series alone.

\subsection{Hidden scale symmetry origin of the method of multiple scales}\label{sec:MMS}

In the method of multiple scales (MMS), the dependent variable is assumed to depend on extra slow scales $\tau_i=\epsilon^i t$, where $i>0$ are guessed, and $\tau_i$ are treated as independent variables. Because at zeroth order the perturbation equation is unchanged by introducing these additional scales, its solution is identical to the naive reference solution but with integration constants replaced by functions of these additional scales. Their functional form is subsequently chosen to ensure the secular terms at higher order vanish~\cite{Hinch1991,Bender1999I}.

To illustrate the relationship between hidden scale symmetries and the method of multiple scales we consider the Mathieu equation (treated by MMS on p.\ 560 of ref.~\cite{Bender1999I}):
\begin{equation}\label{MathieuDE}
\ddot{y}+(a+2\epsilon\cos t)y=0.
\end{equation}
It can be shown that for $a\neq n^2/4\ (n=0,1,2,\dots)$, $y(t)=0$ is stable for sufficiently small $\epsilon$~\cite{Bender1999I}. Investigating the stability boundary of solutions in the $(a,\epsilon)$ plane around $a=1/4$ may be done by expanding $a$ as $a(\epsilon)=1/4+a_1\epsilon+a_2\epsilon^2+\dots$. Expanding $y$ similarly as $y=y^{(0)}+\epsilon y^{(1)}$ yields the perturbation equations to first order:
\begin{equation}
\ddot{y}^{(0)}+\frac{1}{4}y^{(0)}=0,\quad \ddot{y}^{(1)}+(a_1+2\cos t)y^{(1)}=0.
\end{equation}
Solving these yields the first-order perturbation series:
\begin{multline}\label{Mathieupert}
y_1(t)=R\cos(t/2+\theta)+\epsilon A\cos(t/2+\theta)+\epsilon B\sin(t/2+\theta)\\+\epsilon R\left[\frac{1}{2}\cos(3t/2+\theta)+\sin(2\theta)t\cos(t/2+\theta)\vphantom{\frac{3}{2}}-(\cos(2\theta)+a_1)t\sin(t/2+\theta)\right],
\end{multline}
where $A$ and $B$ are integration constants. 

In this example painting is essential to obtain FT equations simple enough to solve. The $t=0$ solution is a function of $\cos\theta$ and $\sin\theta$, but the full solution contains non-divergent terms proportional to $\cos(t/2+\theta)$ and $\cos(3t/2+\theta)$. A $t-\theta$ approximate symmetry converting the former into the latter would be very complex and likely non-analytic. Indeed, explicit computation of such a symmetry proves not to be possible in this case.

Painting Eq.~\eqref{Mathieupert} by replacing only divergent instances of $t$ with a new variable $\mu$ yields:
\begin{multline}
y_1(t)=R\cos(t/2+\theta)+\epsilon A\cos(t/2+\theta)+\epsilon B\sin(t/2+\theta)\\+\epsilon R\left[\frac{1}{2}\cos(3t/2+\theta)+\sin(2\theta)\mu\cos(t/2+\theta)\vphantom{\frac{3}{2}}-(\cos(2\theta)+a_1)\mu\sin(t/2+\theta)\right].
\end{multline}
Now, the finite transformation (FT) equation can be applied for independent variable $\mu$ instead of $t$. (See Appendix~S2 for why painting additionally some but not all convergent terms would have been inconsistent.) This recovers the same equations as are obtained by CGO RG in ref.~\cite{Chen1996}, but time-inverted as expected due to the different direction of integration:
\begin{subequations}
	\begin{align}
	O(\epsilon)&=R'\cos(t/2+\theta)-R\sin(t/2+\theta)\theta'\\
	\therefore O(\epsilon^2)&=R'\cos(t/2+\theta)-R\sin(t/2+\theta)\theta'\nonumber\\&\quad+\epsilon R\left[\sin(2\theta)\cos(t/2+\theta)-(\cos(2\theta)+a_1)\sin(t/2+\theta)\right]\\
	&=R'+\epsilon R\sin(2\theta)\quad=\theta'+\epsilon (\cos(2\theta)+a_1).
	\end{align}
\end{subequations}
The solutions to these FT equations (Eqs~(2.16)-(2.18) from ref.~\cite{Chen1996}) are plotted alongside numerical solutions to Eq.~\eqref{MathieuDE} in Fig.~S1. 

That CGO RG finds the same scales as multiple-scale analysis was discussed in refs~\cite{Chen1994,Chen1996}. It should now be clear that the correspondence in outcomes between these two techniques follows, like their validity, from their mathematical underpinning in hidden scale symmetries. CGO RG, MMS, and the hidden scale symmetry method all find a globally convergent solution in the form of the convergent terms of the perturbation series with the zeroth-order integration constants replaced by slowly varying functions of the independent variables. The hidden scale symmetry method ensures this convergence rigorously and algorithmically by determining these functions through a globally valid symmetry transformation. 
In MMS, by contrast, guesswork is needed to identify the slow scales upon which these functions depend.

\subsection{Lie symmetry interpretation of the Poincare-Lindstedt method}\label{sec:PLM}

The method of strained coordinates, or Poincare-Lindstedt method (PLM), replaces independent variable $t$ with $Q(\varepsilon)t$, where $Q=1+q_1\varepsilon+q_2\varepsilon^2+\dots$. Similarly to MMS, the $q_i$ are then chosen so that secular terms in the perturbation series $y_k=\sum_{i=1}^k\epsilon^iy^{(i)}$ vanish. From the Lie symmetry viewpoint, this is equivalent to introducing a switching parameter $\mu$ in front of the divergent terms in the perturbation series, where $\mu$ can run from 0 to 1, alongside the introduction of $Q$ as outlined above, seeking an approximate symmetry connecting $Q$ and $\mu$, and expanding the resultant finite transformation to finite order in $\varepsilon$. 

This viewpoint allows us to understand when PLM will succeed. Clearly, the perturbation series with arbitrary $Q$ must still be a valid perturbative solution, which (since integration constants $A_i$ can be chosen arbitrarily) in turn requires $y_k(t,\{A_i\})=y_k(f(t,A_1),\{A_{i>1}\})=y_k(f(Qt,\tilde{A}_1),\{A_{i>1}\})$. In this case, PLM just looks for the finite transformation of a hidden scale symmetry involving only $A_1$, and will succeed only when such a symmetry exists.

As an example, we look at the travelling wave solutions of the Korteweg-de Vries (KdV) equation investigated in ref.~\cite{Kevorkian1996} by PLM. A version of the KdV equation is:
\begin{equation}\label{KdVPDE}
u_t+u_x+\frac{\delta^2}{6}u_{xxx}+\varepsilon\frac{3}{2}uu_x=0.
\end{equation}
Looking for a traveling wave solution $u=W(\theta)\equiv W(kx-\omega t)$ reduces this to an ODE:
\begin{equation}\label{KdVODE}
(k-\omega) W'+\frac{\delta^2k^3}{6}W'''+\varepsilon k\frac{3}{2}WW'=0.
\end{equation}
Expanding $W=W^{(0)}+\epsilon W^{(1)}+\epsilon^2 W^{(2)}+O(\epsilon^3)$, the reference solution $W_0=W^{(0)}$ is obtained by solving:
\begin{equation}
(k-\omega) (W^{(0)})'+\frac{\delta^2k^3}{6}(W^{(0)})'''=0,
\end{equation}
yielding:
\begin{equation}\label{refsoln}
	W_0=R\sin(\lambda\theta+\phi)+c_0,\quad \lambda^2=\frac{6(k-\omega)}{\delta^2k^3}.
\end{equation}
To restrict our attention to zero-average solutions, we set $c_0=0$. Next, WLOG we can set $\lambda=1$, as in ref.~\cite{Kevorkian1996}, since $\lambda$ just multiplies the arbitrary $\omega$ and $k$ in Eq.~\eqref{refsoln}. In SI Sec.~S5~A we calculate the terms in the resultant perturbation series to second order, yielding:
\begin{equation}\label{barePLM}
W_2=R\sin(\theta+\phi)-\varepsilon\frac{6 R^2\sin\!\left(\!\frac{\theta}{2}\!\right)\!^3 \sin\!\left(\!\frac{\theta}{2}+2\phi\!\right)}{k^2\delta^2}-\varepsilon^2\frac{27R^3\theta(6\cos(2\phi)-1)\cos(\theta+\phi)}{16k^4\delta^4}+\varepsilon^2\mathcal{R},
\end{equation}
where $\mathcal{R}$ consists of all the non-divergent functions in the second-order term (and is given by Eq.~S32). 

\begin{figure}
	\centering
	\includegraphics[width=\textwidth]{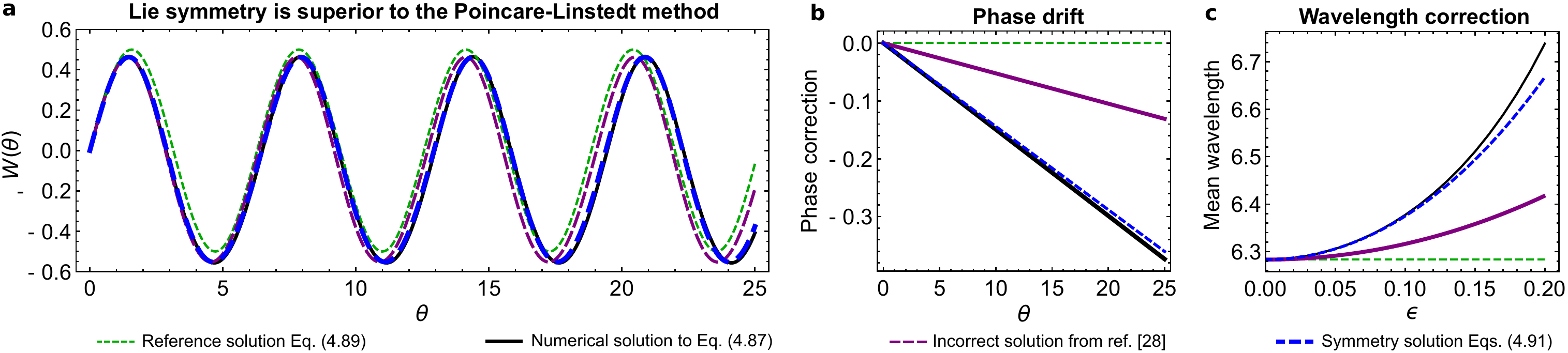}
	\caption{The method of strained coordinates is not always applicable. \textbf{a}: Travelling wave solutions of the Korteweg-de Vries equation satisfy Eq.~\eqref{KdVODE}, which admits a singular perturbation series. This was solved incorrectly by the Poincare-Lindstedt Method in ref.~\cite{Kevorkian1996}; here, it is easily solved correctly by the hidden scale symmetry method ($\epsilon=0.14,\ W(0)=W''(0)=0,\ W'(0)=0.5$). \textbf{b}: The error in the solution from ref.~\cite{Kevorkian1996} becomes more apparent at larger $\theta$, because \textbf{c}: the source of the error is an incorrect modification to the wavelength. The second-order hidden scale symmetry solution holds the correct wavelength up until $\epsilon\simeq 0.15$.}
	\label{Fig5}
\end{figure}

Now, $W_2$ does not satisfy $W_2(\theta,\phi,R)=W_2(f(\theta,\phi),R)$ (and nor does the first order series $W_1$). This implies that PLM does not apply here, and that the solution in the textbook may therefore be incorrect. To verify this, we first compute the correct solution using hidden scale symmetries. Using initial conditions $W(0)=W''(0)=0,\ W'(0)=A_1$ this yields (see SI Sec.~S5~B):
\begin{subequations}
	\begin{equation}
	W(\theta)=A_1\sin((1-q)\theta)+\frac{3\varepsilon A_1^2}{4k^2\delta^2}\left[4\cos{((1-q)\theta)}-\cos{(2(1-q)\theta)}-3\right]
	\end{equation}\vspace{-0.5cm}
	\begin{equation}
	q=\frac{135A_1^2\varepsilon^2}{16k^4\delta^4}.
	\end{equation}
\end{subequations}
In the textbook~\cite{Kevorkian1996} PLM is implemented in a surprising way, in which only $t$ is modified, not $\theta$. This is mathematically equivalent thanks to the arbitrariness of $\omega$, and may be re-expressed as a modification for $\theta$ by calculating how it modifies $\lambda$ from 1 (see SI Sec.~S5~C). Adapting for the initial conditions $W(0)=W''(0)=0,\ W'(0)=A_1$ yields the same expression for $W$, but with $q$ replaced by:
\begin{equation}
q_\text{text}=\frac{27A_1^2\varepsilon^2}{16k^5\delta^4},
\end{equation}
i.e.\ differing to our solution by a factor of $1/(5k)$ in $q$. Plotting against the numerical solution confirms the solution in the textbook is incorrect whereas our solution is correct (Fig.~\ref{Fig5}).

So, although superficially PLM may appear applicable to this problem, a Lie symmetry analysis reveals this not to be the case, and hidden scale symmetry analysis further reveals the solution derived by PLM to be incorrect.

\subsection{Other symmetries can sometimes be exploited when hidden scale symmetries are intractable}\label{sec:pertsym}
Sometimes these methods yield an insoluble RG or FT equation. An example of where this occurs is the modified inviscid Burgers equation as an initial-value problem:
\begin{equation}\label{Hopfmod}
u_t+\epsilon u u_x^2=0,\quad u(t_0,x)=U(t_0,x),
\end{equation}
where $U(t_0,x)$ is an arbitrary initial value. (We can later impose a known initial value at $t_0=0$.) Expanding $u$ in $\epsilon$ as $u=u^{(0)}+\epsilon u^{(1)}+O(\epsilon^2)$ yields the perturbation equations:
\begin{equation}
u^{(0)}_t=0,\quad u^{(1)}_t+u^{(0)} \left(u^{(0)}_x\right)^2=0.
\end{equation}
Assuming $U=O(1)$, the boundary conditions are $u^{(0)}(t_0,x)=U(t_0,x),\ u^{(1)}(t_0,x)=0$. To first order, the perturbation series arising from these is:
\begin{equation}\label{Hopfmodpertsoln}
u_1(t,x,t_0)=U-\epsilon(t-t_0)U U_x^2,
\end{equation}
which is clearly singular. The RG equation resulting from applying Eq.~\eqref{CGOFT} is:
\begin{equation}
U_{t_0}+\epsilon U U_x^2=0,
\end{equation}
but this is just the original DE. Since the perturbation equations are first order the direct hidden scales method will yield an equivalent equation. This shows that the hidden scale symmetry to $O(\epsilon^1)$ is exact, and thus of no help for an approximate solution.

\begin{figure}
	\centering
	\includegraphics[width=0.92\textwidth]{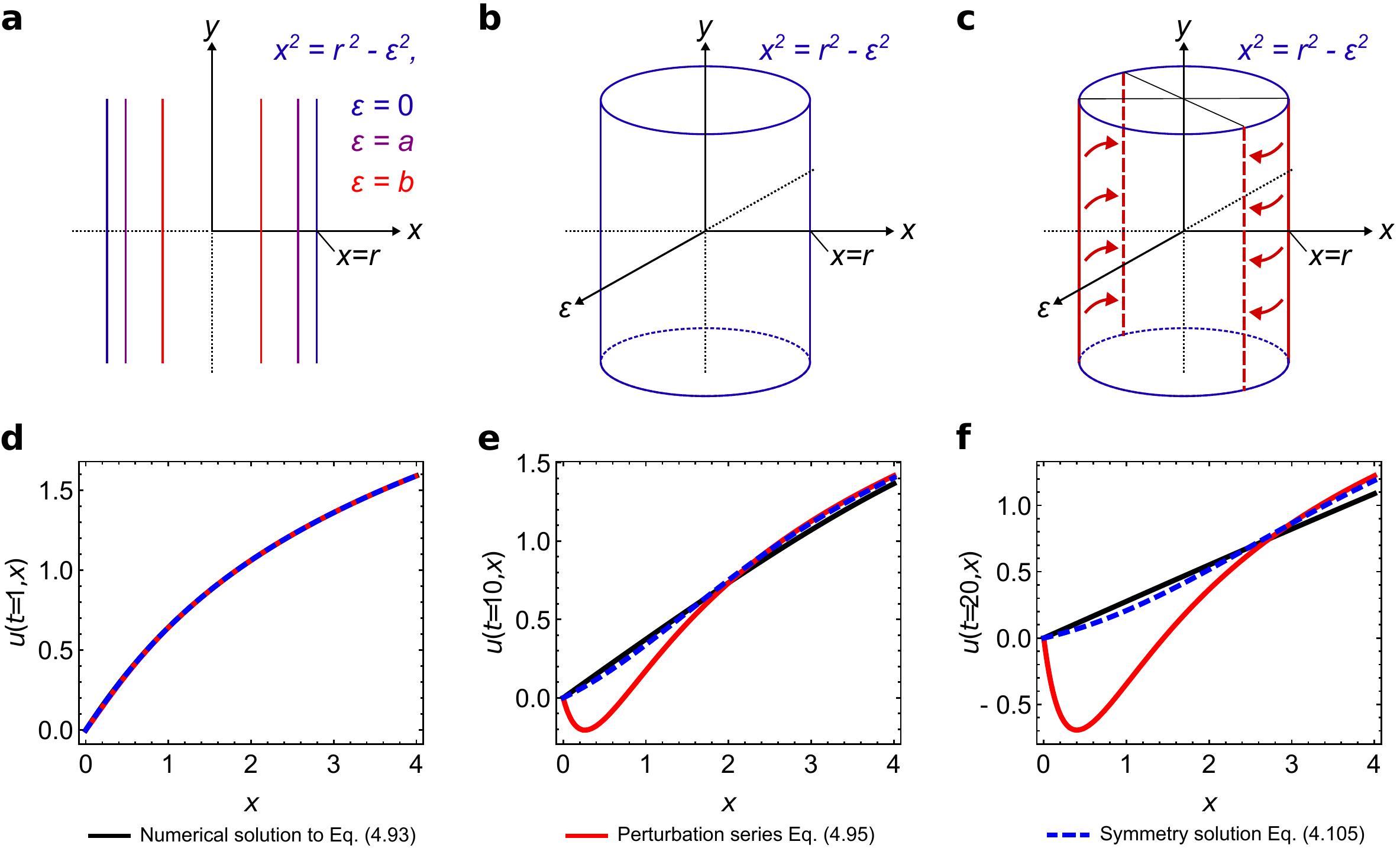}
	\caption{Treating perturbation parameters as variables leads to a richer set of symmetries. \textbf{a}: Two parallel lines with a perturbation in their distance from the origin have no Lie symmetries. \textbf{b}: Treating the perturbation parameter as a new variable turns these lines into a cylinder, which has a rotational Lie point symmetry. \textbf{c}: Any $\epsilon\neq 0$ solution may now be obtained by finite transformation of the $\epsilon= 0$ solution using this symmetry. \textbf{d}-\textbf{f}: Black: Numerical solution of Eq.~\eqref{Hopfmod}. Red: divergent first-order perturbative solution (Eq.~\eqref{Hopfmodpertsoln}). Blue: first-order approximate manifold generated by perturbation symmetry (Eq.~\eqref{Hopfmodsymsoln}). Solutions evaluated at $t=1,\ 10$ and $20$ respectively.}
	\label{Fig7}
\end{figure}

We can instead attempt to solve using another kind of approximate Lie solution symmetry. Extending the manifold to include the perturbation parameter $\epsilon$, and treating it just like an independent variable (Fig.~\ref{Fig7}\textbf{a}-\textbf{b}), permits us to define ``perturbation symmetries'', that act directly on $\epsilon$. If such a symmetry can be found, then by constructing the associated finite transformation it is possible to directly convert the unperturbed solution to the general solution to the full differential equation (Fig.~\ref{Fig7}\textbf{c} shows a geometric analogy). 

Exact perturbation symmetries, computed from the DE directly, were first proposed by Kovalev \textit{et al}~\cite{Kovalev1998}, and this method was demonstrated to correctly solve various DEs that can also be solved by standard methods. They dubbed them ``renormgroup symmetries'' owing to a loose analogy to the CGO RG method. However, as the present paper makes clear, these are distinct from the symmetries that underpin CGO RG, prompting our introduction of the alternative nomenclature ``perturbation symmetries'' to avoid potential confusion. 

Exact perturbation symmetries, since they must be computed from the DEs using traditional Lie symmetry analysis techniques, are out of the scope of the present paper. (In any case, we are not aware of any DE that can be solved by them but that cannot be solved more easily by other methods.) Instead, we are interested in \textit{approximate} perturbation symmetries, that leave the solution manifold invariant to $O(\epsilon^k)$ for some specified order $k$. Symmetries of this type, albeit calculated laboriously in the traditional way from DEs, were proposed in ref.~\cite{Iwasa2006} as a perturbation-series-free alternative to CGO RG, although it was not suggested that it could succeed when CGO RG fails. In this paper we have instead taken the opposite approach, that the availability of a perturbation series enables the straightforward computation of these and other kinds of approximate symmetries that would otherwise be extremely laborious or impossible to determine using traditional methods.

One might expect the generator for a 0\textsuperscript{th} order approximate perturbation symmetry to take the form:
\begin{equation}
\bm{X}^{(0)}=\sum_{i=1}^n\xi_i^{(0)}\frac{\partial}{\partial x_i}+\sum_{j=1}^m\eta_j^{(0)}\frac{\partial}{\partial y_j}+\sum_{i=1}^p\xi_{c_i}^{(0)}\frac{\partial}{\partial c_i}+\xi_\epsilon^{(0)}\frac{\partial}{\partial \epsilon},
\end{equation}
where $\epsilon$ is the perturbation parameter, and $c_i$ are other constant parameters to which the manifold has been extended. (This approach was taken in ref.~\cite{Iwasa2006}.) However, the derivative with respect to $\epsilon$ mixes orders and can be viewed as an $O(\epsilon^{-1})$ operator, inappropriately pulling higher-order terms into the zeroth-order symmetry. The zeroth-order solution to the perturbation problem is the unperturbed solution, and so the zeroth-order symmetries should also be the unperturbed symmetries, which will not be true with the above formulation. Instead, $\epsilon$ should be replaced everywhere in the perturbation series with $\epsilon s$, where $s$ plays the role of the perturbative switching, running from 0 to 1, and $\epsilon$ retains the scale. Seeking approximate perturbation symmetries instead in $s$ then avoids issues of mixing orders. The $k$\textsuperscript{th}-order term in the generator then takes the form:
\begin{equation}\label{genpertsym}
\bm{X}^{(k)}=\sum_{i=1}^n\xi_i^{(k)}\frac{\partial}{\partial x_i}+\sum_{j=1}^m\eta_j^{(k)}\frac{\partial}{\partial y_j}+\sum_{i=1}^p\xi_{c_i}^{(k)}\frac{\partial}{\partial c_i}+\xi_s^{(k)}\frac{\partial}{\partial s}.
\end{equation}
Similarly, if $c_i$ are perturbative, they can be replaced with $\epsilon c_i$ in the DEs, so that $c_i$ are now $O(1)$, and now Eq.~\eqref{genpertsym} no longer mixes orders.

We are now in a position to calculate $O(\epsilon)$ approximate perturbation symmetries for Eq.~\eqref{Hopfmod}. Since we are not calculating envelope symmetries we no longer need $t_0$ and can set it to the value for which we know $U$, say $t_0=0$. We first consider ones that act only on $x$ and $\epsilon$. Such symmetries do not act on $u$ directly, so should satisfy the following equation to $O(\epsilon^2)$:
\begin{equation}
(\bm{X}^{(0)}+\epsilon\bm{X}^{(1)})(U-\epsilon s tU U_x^2)|_{u=U-\epsilon s tU U_x^2}=0.
\end{equation}
At zeroth order this gives $\xi_x^{(0)}=0$; at first order, we find:
\begin{equation}
	(\xi_x^{(1)}U_x-\xi_s^{(0)}t U U_x^2)|_{U=u}=0\quad \Rightarrow \xi_x^{(1)}=t u U_x\xi_s^{(0)}.
\end{equation}
Thus, the perturbation symmetry is:
\begin{equation}
\bm{X}(x,s)=\xi_s^{(0)}\left(\frac{\partial}{\partial s}+\epsilon t u U_x\frac{\partial}{\partial x}\right).
\end{equation}
Now let $x=H(u)$ be the inverse form of the equation $u=U(x)$. Since $s$ can conveniently be chosen to parametrize it, the corresponding finite transformation is:
\begin{equation}
\frac{d\tilde{x}}{ds}=\epsilon tu U_{\tilde{x}}\ \ \Rightarrow \int_{H(u)}^{x}\frac{d\tilde{x}}{U_{\tilde{x}}}=\epsilon [s]_0^1 tu=\epsilon tu,
\end{equation}
where the limits in the integral follow because when $s=0,\ u=U(x)$, and therefore $x=H(u)$.

To perform the integral an explicit initial value $U$ is required, whereupon the resultant equation can be solved for $u$ to give a globally valid non-divergent approximate solution. As an example, we consider the initial value $U(x)=\ln(1+x)$, in which case the perturbation series becomes:
\begin{equation}
u=\ln(1+x)-\epsilon t\frac{\ln(1+x)}{(1+x)^2}+O(\epsilon^2),
\end{equation}
and, with $H(u)=e^u-1$, the integral becomes:
\begin{equation}
\epsilon tu=\left[\frac{1}{2}\tilde{x}^2+\tilde{x}\right]_{e^u-1}^x\!\!\!\!=\frac{1}{2}x^2+x+1-e^u-(e^u-1)^2.
\end{equation}
Some rearrangement gives the approximate symmetry solution as:
\begin{equation}\label{Hopfmodsymsoln}
u(t,x)=\frac{1}{2\epsilon t}(x+1)^2-\frac{1}{2}W\left[\frac{1}{\epsilon t}e^{(x+1)^2/(\epsilon t)}\right]+O(\epsilon^2),
\end{equation}
where $W[\dots]$ is the Lambert W-function or the product logarithm. In Fig.~\ref{Fig7}\textbf{d}-\textbf{f} we demonstrate that this is indeed a globally valid accurate approximate solution, and also confirm the divergence of the perturbation series.

\section{Discussion}\label{sec:conclusions}

The ideas developed in this paper raise a variety of further questions, not all of which are directly related to the key theme of the paper, but which are nevertheless interesting. We discuss several of these here, and outline potential future research directions to investigate them in more detail.

\subsection{A potential for unification of approximate methods}
Using examples we have demonstrated the basis in approximate Lie symmetries of the method of multiple scales, the Poincare-Lindstedt method, CGO RG and RG/E. All approaches indirectly exploit a class of symmetries that we have coined ``hidden scale symmetries''; all but RG/E involve differing degrees of trial-and-error to do so. We have also shown that these symmetries may instead be directly identified and integrated to solve singular perturbation problems by using an explicitly Lie theoretic approach.

We have also shown that other kinds of approximate symmetries can be identified and exploited when hidden scale symmetries are not obtainable. We suspect that most singular perturbation methods not examined in this paper must also depend in some way on various classes of approximate Lie symmetries. Their explicit reformulation in terms of symmetry transformations may provide technical and conceptual advantages, as we have demonstrated for the methods studied in this paper. Moreover, perhaps additional methods might be developed based on exploiting new classes of approximate symmetry not heretofore used. Also, potential links between approximate Lie symmetry and nominally nonperturbative approximate methods such as invariant manifold theory~\cite{Roberts2020} could be investigated in greater depth.

\subsection{Renormalization group reformulated as approximate Lie symmetry}

CGO RG contains coarse-graining/splitting and renormalizing steps because it was developed as an analogy of perturbative RG. Our interpretation for the success of the CGO RG method in solving singular perturbation problems is its identification and exploitation of a class of approximate Lie symmetries of the solution that we have termed ``hidden scale symmetries.'' When viewed in light of symmetry transformations the splitting and renormalizing steps are revealed as mostly redundant; their removal simplifies the calculations, particularly when multiple splittings are possible. 

Reversing the analogy that was the original inspiration for this technique, this suggests that under certain circumstances, perturbative renormalization group in momentum space may be better understood as an approximate Lie symmetry of the integral studied, hearkening back to the original links between exact renormalization group and traditional Lie theory \cite{bogoliubov}. In particular, this interpretation can hold only if information is not lost, i.e.\ the RG transformation is reversible. So, the ``coarse-graining'' approach to RG of Kadanoff does not have a straightforward interpretation in terms of approximate Lie symmetries.

On a related note, we would argue that the equivalence to approximate Lie solution symmetries of the ``envelopes'' at the heart of Kunihiro's geometric formulation of RG provides an intuitive interpretation of the latter concept.

\subsection{Broader applications of our method for calculating symmetries}
Our results on singular perturbation theory would not have been possible without developing a new method for calculating approximate Lie symmetries to the solutions of DEs. Any such symmetry that can be calculated via the DE using standard methods can be calculated more easily using our approach (e.g.\ the perturbation symmetry of the underdamped harmonic oscillator; see SI Sec.~S6). This may have applications in the field of Lie symmetries extending beyond the present paper. 

Symmetries involving the integration constants $A_i$ can \textit{only} be calculated by our approach. Since $A_i$ remain unfixed, and change implicitly with the independent variables $x$, the method can only precisely calculate such symmetries when they act on $A_i$ and $x$ directly, like hidden scale symmetries do. If the integration constants enter the generator but the symmetry does not act on them directly, additional approximation steps must be taken. For instance, using appropriate trial functions for the integration constants (as was effectively done in Eq.~\eqref{Hopfmodpertsoln}) turns the solution-by-symmetry method into a kind of self-consistent approximate method. Such methods should be the subject of a dedicated future study. Furthermore, our method should in principle find the approximate Lie solution symmetries of other kinds of equations such as algebraic or integral equations; given the great power of these symmetries, we therefore expect that this method may find applications unrelated to DEs in the future.

\subsection{Advantages and disadvantages of the hidden-scale approach and MMS}

Both the hidden scale method and CGO RG share the advantage over MMS that timescales emerge automatically and need not be guessed. Sometimes, to determine the correct timescales using MMS, it is necessary to calculate the perturbation series to inconveniently high order. For instance, in Eq.~\eqref{MathieuDE}, the fourth-order perturbation series is needed to solve to second order by MMS~\cite{Chen1996}. On the other hand, sometimes the reverse is true and to get an $O(\varepsilon^n)$ global approximate solution using MMS it is only necessary to compute the perturbation series to $O(\varepsilon^{n-1})$ and then simply to eliminate those components from the \textit{equation} for the $O(\varepsilon^n)$ term that will give rise to divergent components in its solution. With CGO RG and hidden scales the perturbation series must generally be computed to the same order as that of the desired solution. When the perturbation series is challenging or time-consuming to compute, e.g.\ for certain partial differential equations, and MMS requires only an $O(\varepsilon^{n-1})$ series, MMS may be the preferred method, despite the need to guess the timescales.

PLM has been presented as a more primitive version of MMS, out of which MMS evolved~\cite{Kevorkian1996}; however, there is a fundamental difference. We found that whereas MMS always has a mathematically rigorous basis in hidden scale symmetries, PLM is a viable method only when its implementation serendipitously mirrors that of MMS. So PLM succeeds when it unconsciously exploits hidden scale symmetries, and fails when it attempts to exploit transformations that are not valid symmetries.

Overall, though, we believe the direct hidden scales symmetry method will most often be the best choice. We have demonstrated that it can straightforwardly and algorithmically solve problems for which earlier techniques are impractical or require significant guesswork. Our hope is that its greater simplicity, intuitiveness, and lack of operational ambiguity might encourage more routine attempts to approximately solve differential equations analytically in the future, and as before, continue to complement numerical methods of analysis.\vskip6pt

\begin{acknowledgments}
	We acknowledge support from the Lindemann Trust Fellowship, English-Speaking Union (AJD), the MacArthur Foundation (LM), the Simons Foundation (LM) and the Henri Seydoux Fund (LM). We are grateful to Teiji Kunihiro and to Nigel Goldenfeld for providing invaluable feedback on the preprint.
\end{acknowledgments}


\pagebreak

\part*{Supporting Information: Approximate Lie symmetries and singular perturbation theory}
\setcounter{section}{0}
\setcounter{equation}{0}
\setcounter{figure}{0}
\def\theequation{S\arabic{equation}}
\def\thesection{S\arabic{section}}

\section{Overview of some relevant established methods}\label{sec:methods}

\subsection{Regular and singular perturbations of DEs}

A wide class of DEs $F=F_0+\epsilon F_1$, known as ``perturbed DEs'', may be expressed as the sum of a solvable differential equation $F_0=0$ (the ``reference problem'') and another term $F_1$. The latter is multiplied by a switching parameter $\epsilon$, which may or may not be small, and the dependent variable $y$ is then expanded in this parameter, defining a ``perturbation series'' $\sum_{i=0}^\infty\epsilon^iy^{(i)}$. This leads to a set of simpler DEs, one for each term in the series expansion, with $y_0$ simply being the solution to $F_0=0$. These DEs are generally simpler to solve than the original perturbed DE, at least for the lower order terms in the expansion, yielding an approximate solution to the desired order. This framework generalizes naturally to DEs with multiple terms $F_i$ added to the reference system, multiplied by switching parameters $\epsilon_i$.

The key limitation of this technique is that many perturbation problems are ``singular'', defined as one whose perturbation series has a vanishing radius of convergence, so is \textit{not} a convergent power series in $\epsilon$~\cite{WikiSingPert,Bender1999I}. Crucially, this means the exact solution even for $\epsilon\to 0$ must be qualitatively different from the $\epsilon=0$ solution to the unperturbed problem. Conversely, if the perturbation series has a finite radius of convergence within the region of phase space of interest, it is known as a ``regular'' perturbation problem.

In order to obtain a valid perturbation series, it is necessary that the order of $F_0$ is the same as that of $F$. Otherwise, the solution to $F_0$ will not have enough integration constants to satisfy the boundary or initial conditions. This can usually be ensured by rescaling the dependent and independent variables appropriately so that $\epsilon$ no longer multiplies the highest-order derivatives, called transforming into the ``inner layer''.

\subsection{Chen-Goldenfeld-Oono Renormalization Group (CGO RG) and Kunihiro's formulation (RG/E)}\label{methods:CGO}
The CGO RG method uses as a starting point an $O(\epsilon^k)$ singular perturbation series $y_k=\sum_{i=1}^k\epsilon^iy^{(i)}$ that has been transformed into the inner layer, is divergent in an independent variable $x$, and whose constants of integration $A_j$ have not yet all been chosen to match the initial or boundary conditions. In its original formulation~\cite{Chen1994}, the idea was to ``split'' the independent variable by rewriting it as e.g.\ $(x-\mu)+\mu$, delete the terms in $\mu$ and assume they have been ``absorbed'' into the now $\mu$-dependent constants of integration $A_j$, in a kind of direct analogy with perturbative RG from high-energy and condensed matter physics. After following this procedure $y_k$ is now said to be ``renormalized''. The $t_0$ dependence is finally determined by insisting $y_k$ satisfy the ``RG equation'':
\begin{equation}\label{pertflow}
\frac{dy_k(t,\mu)}{d\mu}\bigg|_{\mu=x}=\frac{\partial y_k}{\partial \mu}\bigg|_{\mu=x}+\sum_j\frac{\partial y_k}{\partial A_j}\frac{dA_j}{d\mu}\bigg|_{\mu=x}=0.
\end{equation}

Essentially any splitting $f(x)\to (f(x)-f(\mu))+f(\mu)$ was permitted in this original formulation of CGO RG. However, Kunihiro developed an alternative RG-free geometric formulation based on the idea that the CGO RG equation finds the ``envelope'' of a family of functions (the curve that is tangent to all its members)~\cite{Kunihiro1995,Kunihiro1997,Kunihiro1997b}. These functions are defined by the perturbation series solved for initial or boundary conditions imposed at different arbitrary values $x_0$. It was proved~\cite{Kunihiro1995,Kunihiro1997} that the envelope of the $n$\textsuperscript{th} order perturbation series calculated from different initial or boundary conditions at $x=x_0$ yields a globally valid $O(\epsilon^n)$ approximate solution to the original equations. An additional valuable perspective was later provided by Ei, Kunihiro \textit{et al}~\cite{Ei2000,Kunihiro2006}, describing the construction of invariant manifolds by the RG equation. See \cite{Kunihiro_book} for a unified account of all these contributions.

The splitting and renormalizing steps of CGO RG are effectively replaced in Kunihiro's formulation (termed RG/E) by the single step of using an arbitrary $x_0$ and redefining the integration constants $A_i$ such that the divergent terms vanish at $x=x_0$. RG/E and CGO RG become equivalent in practice when the splitting in the latter has been introduced in such a way that relabelling $\mu=x_0$ in the renormalized expression yields a valid perturbation series of this type. RG/E can thus alternatively be viewed as CGO RG with a reduced number of permissible splittings. Despite these differences, both CGO RG and RG/E have seen widespread use since their introduction.

\section{Painting}\label{app_painting}
In the main text the concept of ``painting'' was introduced, whereby only divergent instances of the independent variable $x$ are replaced by new variable $\mu$. Subsequently, symmetries connecting the integration constants with $\mu$ are sought, and integrated from $\mu=0$ to $\mu=x$ to yield a global approximate solution. Painting is always mathematically valid, and the resultant global solution is convergent by construction if all divergent terms have been painted in both the original perturbative solution and its derivatives up to the order below that of the original DE. Its motivation is two-fold. First, finite transformation equations of the unpainted symmetry are sometimes too complicated to solve analytically. Painting and then requiring independence from $x$ can often yield simpler hidden scale symmetry generators, whose finite transformation equations are easier to solve. This is the case in the example studied in Sec.~\textbf{4(i)} in the main text, for instance.

Second, by requiring independence of the generator from $x$, the number of derivatives of the perturbation series that need be calculated and used in the symmetry calculation to fully constrain the generator is reduced. This is seen in the examples studied in both Secs.~\textbf{4(d)} and \textbf{4(i)} in the main text: requiring invariance of solely the perturbative solution and insisting on $x$-independence yields two independent FT equations. The mathematical justification for this is a little more involved. The symmetry calculated in this way from the perturbation series and possibly some of the lower-order derivatives must also leave the higher-order derivatives invariant, where divergent terms in the higher order derivatives have also been painted. If it does not, then it is not a true hidden scale symmetry of the exact solution, and no $x$-independent $A-\mu$ symmetry can exist for this choice of painting. How to ensure that the higher order derivatives are indeed invariant, without explicitly calculating them? Here we rely on consistency of timescales. If only divergent terms are painted, and all divergent terms are painted, then only the hidden (slow) scale is painted, and painted in its entirety. Taking the derivative of the perturbation series or its derivatives does not mix the slow and fast scales. So any symmetry that acts only on the hidden scale in the lower-order derivatives can act only on the hidden scale in the higher derivatives too, and must necessarily leave it invariant.

\section{Summing the most divergent terms at each order in the perturbative solution to Eq.~(\ref{badworseterrible})}\label{app_infinitesum}
We first calculate the most divergent term at each order in the perturbation expansion in $a$, up to arbitrary order. Using subscripts rather than superscripts to indicate orders for convenience here, the $n$\textsuperscript{th} order perturbation equation is:
\begin{equation}\label{nthorder}
u_n''+\left(1+\frac{1}{x}\right)u_n'=-\sum_{i=1}^{n-1}u_i'u_{n-i}'-\sum_{i=1}^{n-1}u_iu_{n-i}'.
\end{equation}
If we guess that the most divergent $n$\textsuperscript{th} order term is $\bar{u}_n=(-1)^{n+1}c_nB^ne_1(x)^n$ for any $n$, then we have:
\begin{align}
\bar{u}_n'&=n(-1)^{n}c_nB^ne_1(x)^{n-1}\frac{e^{-x}}{x}\\
\bar{u}_n''&=n(-1)^{n+1}c_nB^ne_1(x)^{n-1}\left(\frac{e^{-x}}{x}+\frac{e^{-x}}{x^2}\right)\nonumber\\
&\quad+n(n-1)(-1)^{n+1}c_nB^ne_1(x)^{n-2}\left(\frac{e^{-x}}{x}\right)^2.
\end{align}
Substituting in, the most divergent terms in the LHS become:
\begin{multline}
-n(-1)^{n}c_nB^ne_1(x)^{n-1}\left(\frac{e^{-x}}{x}+\frac{e^{-x}}{x^2}\right)
-n(n-1)(-1)^{n}c_nB^ne_1(x)^{n-2}\left(\frac{e^{-x}}{x}\right)^2\\
+\left(1+\frac{1}{x}\right)n(-1)^{n}c_nB^ne_1(x)^{n-1}\frac{e^{-x}}{x}=-n(n-1)(-1)^{n}c_nB^ne_1(x)^{n-2}\left(\frac{e^{-x}}{x}\right)^2.
\end{multline}
We note that $e^{-x}/x$ diverges faster as $x\to 0$ than $e_1(x)$ does; therefore, the first sum in the RHS of Eq.~\eqref{nthorder} diverges faster than the second if our guess as to $\bar{u}$ is correct. So, the most divergent terms in the RHS become:
\begin{multline}
-\sum_{i=1}^{n-1}u_i'u_{n-i}'=-\left(\frac{e^{-x}}{x}\right)^2
\sum_{i=1}^{n-1}i(-1)^{i}c_iB^ie_1(x)^{i-1}(n-i)(-1)^{n-i}c_{n-i}B^{n-i}e_1(x)^{n-i-1}\\
=-\left(\frac{e^{-x}}{x}\right)^2(-1)^{n}B^ne_1(x)^{n-2}\sum_{i=1}^{n-1}ic_i(n-i)c_{n-i}.
\end{multline}
Equating LHS and RHS we are left with:
\begin{equation}
n(n-1)c_n=\sum_{i=1}^{n-1}ic_i(n-i)c_{n-i}.
\end{equation}
This equality holds if $c_n=1/n\ \forall n$. Since it holds for the first two terms in the series, the proof by induction is complete, and we can write the infinite series for the most divergent perturbative terms as:
\begin{equation}
\bar{u}=1+\sum_{n=1}^{\infty}(-1)^{n+1}\frac{1}{n}B^ne_1(x)^n.
\end{equation}
The radius of convergence $a^*$ is given by:
\begin{equation}
1=\lim_{n\to\infty}\left|\frac{\bar{u}_{n+1}}{\bar{u}_{n}}\right|=\lim_{n\to\infty}a^*\left|B\right|\frac{n}{n+1}e_1(x)=a^*\left|B\right|e_1(x).
\end{equation}
If $B$ is chosen to match the boundary condition $u(\epsilon)=1-a$, as in the bare perturbation series, then this becomes:
\begin{equation}
a^*=\frac{e_1(\epsilon)}{e_1(x)}.
\end{equation}
Since the radius of convergence never vanishes, said perturbation series is regular not singular. Moreover, this equals 1 at $x=\epsilon$, and exceeds 1 at any other value of $x$ in the domain of interest. So, we can always set $a=1$ to get a uniformly valid approximate solution to the original DEs.

We also recognize the sum as an expansion of a logarithm, giving:
\begin{equation}
\bar{u}=1+\ln\left(1+aBe_1(x)\right),
\end{equation}
and if we only now satisfy the boundary conditions then the $u(\infty)$ BC is already satisfied, but we also must satisfy:
\begin{align}
\bar{u}(\epsilon)&=1+\ln\left(1+aBe_1(\epsilon)\right)=1-a,\\
\Rightarrow & e^{-a}-1=aBe_1(\epsilon)\\
\therefore & \bar{u}=1+\ln\left(1+(e^{-a}-1)\frac{e_1(x)}{e_1(\epsilon)}\right).
\end{align}
Setting $a=1$ we recover Eq.~~\eqref{terriblesoln2o}.

\section{Imposing the $\rho=R$ boundary condition on the general solution to the Oseen equation}\label{SI_Oseeninner}

For the solution Eq.~\eqref{Oseensemigensoln} to the first-order boundary condition perturbation equation Eq.~\eqref{Oseen2} to satisfy the $\rho=R$ boundary conditions within Eq.~\eqref{BCOseen2}, we require that, $\forall n>1$,
\begin{align}
B_n &R^{-n}+\sum_{m=0}^{\infty}X_m R\,\Phi_{m,n}\!\left(\frac{R}{2}\right)=0.\\
\therefore& -n B_n R^{-n-1}=\sum_{m=0}^{\infty}X_m n\,\Phi_{m,n}\!\left(\frac{R}{2}\right).
\end{align}
We also require that, $\forall n>1$,
\begin{align}
-n B_n &R^{-n-1}+\sum_{m=0}^{\infty}X_m \left[\Phi_{m,n}\!\left(\frac{R}{2}\right)+\frac{R}{2}\,\Phi_{m,n}'\!\left(\frac{R}{2}\right)\right]=0.\\
\therefore& \sum_{m=0}^{\infty}X_m \left[(n+1)\Phi_{m,n}\!\left(\frac{R}{2}\right)+\frac{R}{2}\,\Phi_{m,n}'\!\left(\frac{R}{2}\right)\right]=0.\label{BCresult}
\end{align}
Extending to $n=1$, we have:
\begin{align}
B_1 &R^{-1}+\sum_{m=0}^{\infty}X_m R\Phi_{m,1}\!\left(\frac{R}{2}\right)=-R\\
-B_1 &R^{-2}+\sum_{m=0}^{\infty}X_m \left[\Phi_{m,1}\!\left(\frac{R}{2}\right)+\frac{R}{2}\,\Phi_{m,1}'\!\left(\frac{R}{2}\right)\right]=-1\\
\therefore& 2B_1R^{-1}-\sum_{m=0}^{\infty}X_m\frac{R^2}{2}\Phi_{m,1}'\!\left(\frac{R}{2}\right)=0\\
\therefore& B_1=\sum_{m=0}^{\infty}X_m\frac{R^3}{4}\Phi_{m,1}'\!\left(\frac{R}{2}\right)\\
\Rightarrow& \sum_{m=0}^{\infty}X_m\left[R\Phi_{m,1}\!\left(\frac{R}{2}\right)+\frac{R^2}{4}\,\Phi_{m,1}'\!\left(\frac{R}{2}\right)\right]=-R\\ \Rightarrow&\sum_{m=0}^{\infty}X_m\left[2\Phi_{m,1}\!\left(\frac{R}{2}\right)+\frac{R}{2}\,\Phi_{m,1}'\!\left(\frac{R}{2}\right)\right]=-2.
\end{align}
Pulling this together yields finally Eq.~\eqref{mateq}.

\section{Travelling wave KdV equation}

\subsection{Perturbative solution to second order}\label{app:pertsolnKdV}
To restrict our attention to zero-average solutions, we set $c_0=0$. Next, WLOG we can set $\lambda=1$, as in ref.~\cite{Kevorkian1996}, since $\lambda$ just multiplies the arbitrary $\omega$ and $k$ in Eqs.~\eqref{refsoln}. Setting $c_0=0,\ \lambda=1$ reduces the reference solution to:
\begin{subequations}\label{refsolnred}
	\begin{align}
	W_0&=R\sin(\theta+\phi),\\
	1&=\frac{6(k-\omega)}{\delta^2k^3}.\label{dispersion}
	\end{align}
\end{subequations}
Using the dispersion relation Eq.~\eqref{dispersion}, the travelling wave KdV equation Eq.~\eqref{KdVODE} reduces to:
\begin{align}
0&=\frac{\delta^2k^3}{6} W'+\frac{\delta^2k^3}{6}W'''+\varepsilon k\frac{3}{2}WW'\\
&=W'''+W'+\varepsilon \frac{9}{\delta^2k^2}WW'.\label{KdVODEred}
\end{align}
The first order perturbation equation is then:
\begin{multline}
(W^{(1)})'''+(W^{(1)})'=- \frac{9W_0W_0'}{\delta^2k^2}=- \frac{9R^2}{\delta^2k^2}\sin(\theta+\phi)\cos(\theta+\phi)\\=- \frac{9R^2}{2\delta^2k^2}\sin(2(\theta+\phi)).
\end{multline}
Since the inhomogeneous term is functionally independent from the complementary function (the latter being functionally identical to the reference solution here), we expect no divergent term to arise in the particular integral. Solving the first order perturbation equation in Mathematica with ICs $W^{(1)}(0)=(W^{(1)})'(0)=(W^{(1)})''(0)=0$ to ensure the zeroth order integration constants contain all the information needed to move the solution anywhere then yields:
\begin{equation}
W^{(1)}=-\frac{6 R^2\sin\left(\frac{\theta}{2}\right)^3 \sin\left(\frac{\theta}{2}+2\phi\right)}{k^2\delta^2}.
\end{equation}
As expected, this is not divergent; therefore, we will add this to the reference solution to make the special solution to be symmetry-transformed. The second order perturbation equation is:
\begin{equation}
(W^{(2)})'''+(W^{(2)})'=- \frac{9(W_0(W^{(1)})'+W_1W_0')}{\delta^2k^2}.
\end{equation}
Solving this again in Mathematica with ICs $W^{(2)}(0)=(W^{(2)})'(0)=(W^{(2)})''(0)=0$
The second order term may be written as $W^{(2)}=W^{(2)}_{d}+\mathcal{R}$, where $W^{(2)}_{d}$ consists of the divergent terms, and is given by:
\begin{equation}
W^{(2)}_{d}=-\frac{27R^3\theta(6\cos(2\phi)-1)\cos(\theta+\phi)}{16k^4\delta^4},
\end{equation}
and $\mathcal{R}$ contains only non-divergent terms:
\begin{multline}\label{mathcalR}
\mathcal{R}=-\frac{9R^3}{64k^4\delta^4}\left[28\cos(\phi)\sin(\theta)+57\cos(3\phi)\sin(\theta)\right.\\\left.-54\sin(\theta-\phi)-8\sin(2\theta-\phi)-72\sin(\phi)-8\cos(\theta)\sin(\phi)+24\sin(3\phi)+51\cos(\theta)\sin(3\phi)\right.\\\left.+18\sin(\theta+\phi)+3\sin(3(\theta+\phi))-54\sin(\theta+3\phi)-24\sin(2\theta+3\phi)\right].
\end{multline}

\subsection{Hidden scale FT equations to second order}\label{app:FTeqsKdV}
Painting the divergent instance of $\theta$ in Eq.~\eqref{barePLM} yields:
\begin{multline}
W_2=R\sin(\theta+\phi)-\varepsilon\frac{6 R^2\sin\left(\frac{\theta}{2}\right)^3 \sin\left(\frac{\theta}{2}+2\phi\right)}{k^2\delta^2}\\-\varepsilon^2\mu\frac{27R^3(6\cos(2\phi)-1)\cos(\theta+\phi)}{16k^4\delta^4}+\varepsilon^2\mathcal{R}.
\end{multline}
Differentiating with respect to $\mu$ then yields the FT equations. At zeroth order, these are:
\begin{subequations}
	\begin{align}
	R\cos(\theta+\phi)\phi'+R'\sin(\theta+\phi)&=O(\varepsilon)\\
	\therefore \phi'=R'&=O(\varepsilon).
	\end{align}
\end{subequations}
Thus the derivative with respect to $\mu$ of the first-order term in the perturbation series, which does not contain $\mu$ explicitly, is $O(\varepsilon^2)$. So, in fact, from the derivative of the overall first-order series, $R'=\phi'=O(\varepsilon^2)$. Thus the derivative with respect to $\mu$ of the first-order term in the perturbation series is in fact $O(\varepsilon^3)$ and can be dropped from the derivative of the overall second-order series. This leaves:
\begin{equation}
R\cos(\theta+\phi)\phi'+R'\sin(\theta+\phi)-\varepsilon^2\frac{27R^3[6\cos(2\phi)-1]\cos(\theta+\phi)}{16k^4\delta^4}=O(\varepsilon^3).
\end{equation}
So, to $O(\varepsilon^3)$ we find $R'=0$ so $R(\mu)=\tilde{R}$ (where $\tilde{R}$ is calculated from the available boundary conditions for the problem). Cancelling terms, we are left with:
\begin{equation}
\phi'-\varepsilon^2\frac{27\tilde{R}^2[6\cos(2\phi)-1]}{16k^4\delta^4}=O(\varepsilon^3).
\end{equation}
Since $\phi(\mu)=\tilde{\phi}+O(\varepsilon^2)$, we can simplify this as:
\begin{equation}
\phi'-\varepsilon^2\frac{27\tilde{R}^2[6\cos(2\tilde{\phi})-1]}{16k^4\delta^4}=O(\varepsilon^3).
\end{equation}
Integrating from $\mu=\theta$ to $\mu=0$ with the boundary condition $\phi(\mu=\theta)=\tilde{\phi}$, we have finally the second-order relation:
\begin{equation}\label{finitetransKdV}
\phi-\tilde{\phi}=-\varepsilon^2\theta\frac{27\tilde{R}^2[6\cos(2\tilde{\phi})-1]}{16k^4\delta^4}.
\end{equation}
The special solution is $W_1$ solved instead for arbitrary integration constants:
\begin{equation}
W(\theta)=R\sin(\theta+\phi)+\varepsilon\left[R_1\sin(\theta+\phi)+c_1-\frac{3 R^2}{4k^2\delta^2}\cos{(2(\theta+\phi))}\right].
\end{equation}
We substitute Eq.~\eqref{finitetransKdV} alongside $R=\tilde{R}$, giving:
\begin{subequations}
	\begin{multline}
	W(\theta)=\tilde{R}\sin((1-q)\theta+\tilde{\phi})\\+\varepsilon\left[R_1\sin((1-q)\theta+\tilde{\phi})+c_1\vphantom{\frac{3 \tilde{R}^2}{4k^2\delta^2}}-\frac{3 \tilde{R}^2}{4k^2\delta^2}\cos{(2((1-q)\theta+\tilde{\phi}))}\right].
	\end{multline}\vspace{-0.5cm}
	\begin{equation}
	q=\frac{27\tilde{R}^2[6\cos(2\tilde{\phi})-1]}{16k^4\delta^4}.
	\end{equation}
\end{subequations}
Finally, $\tilde{R},\ \tilde{\phi},\ R_1,\ \phi_1$ and $c_1$ are chosen to match the boundary or initial conditions order-by-order.

\subsection{Converting strained $t$-coordinate into $\theta$-coordinate}\label{app:textconversion}

The solution by strained coordinates in ref.~\cite{Kevorkian1996} is, for our boundary conditions:
\begin{subequations}
	\begin{align}
	\omega t&\to t\left(\omega_0+\epsilon^2\omega_2\right)\\
	\omega_0&=k-k^3\frac{\delta^2}{6}\\
	\omega_2&=\frac{9A_1^2}{16\delta^2k^2},
	\end{align}
\end{subequations}
where $A_1$ is the constant of integration in front of $\cos$ in the reference solution (here, denoted by $R=\tilde{R}$). So, $\lambda$ has changed by:
\begin{align}
\lambda^2&=\frac{6(k-\omega)}{\delta^2k^3}\to \frac{6(k-\omega_0-\epsilon^2\omega_2)}{\delta^2k^3}\\
&=\frac{6\left(k^3\frac{\delta^2}{6}-\epsilon^2\frac{9\tilde{R}^2}{16\delta^2k^2}\right)}{\delta^2k^3}\\
\therefore\lambda&=\left(1-\epsilon^2\frac{27\tilde{R}^2}{8\delta^4k^5}\right)^{1/2}\simeq 1-\frac{27\tilde{R}^2\epsilon^2}{16\delta^4k^5}.
\end{align}

\section{Perturbation symmetries with no dependence on integration constants}\label{app:pertsymnoICs}
True non-asymptotic perturbation symmetries that do not depend on the independent variable can exist only if they do not depend on integration constants (which must depend on the independent variable to be meaningful in a symmetry context). These do indeed exist for some systems; we illustrate using the underdamped harmonic oscillator:
\begin{equation}
\frac{d^2y}{dt^2}+\epsilon\frac{dy}{dt}+y=0.
\end{equation}
To second order, the generator for the perturbation symmetry is:
\begin{equation}\label{2totpertsymgen}
\bm{X}_2=\frac{\partial}{\partial s}+\sum_{k=0}^2\epsilon^k\left(\xi^{(k)}\frac{\partial}{\partial t}+\eta^{(k)}\frac{\partial}{\partial y}\right),
\end{equation}
where W.L.O.G.\ the element of the tangent vector in the direction of $s$ has been set to 1. So, the $k$-th order component of the generator is:
\begin{equation}\label{kpertsymgen}
\bm{X}^{(k)}=\delta_{k,0}\frac{\partial}{\partial s}+\left(\xi^{(k)}\frac{\partial}{\partial t}+\eta^{(k)}\frac{\partial}{\partial y}\right).
\end{equation}

Expanding $y$ perturbatively as $y^{(0)}+\epsilon y^{(1)}+\epsilon^2y^{(2)}$, we have $y^{(0)}=A\sin(t+\theta)$ at zeroth order. At first order:
\begin{align}
\frac{d^2y^{(1)}}{dt^2}+y^{(1)}&=-A\cos(t+\theta)\\
\Rightarrow y^{(1)}(t)&=-\frac{A}{2}t\sin(t+\theta) +C.F.
\end{align}
At second order:
\begin{align}
\frac{d^2y^{(2)}}{dt^2}+y^{(2)}&=\frac{A}{2}t\cos(t+\theta)+\frac{A}{2}\sin(t+\theta)\\
\Rightarrow y^{(2)}(t)&=\frac{A}{8}t^2\sin(t+\theta)-\frac{A}{8}t\cos(t+\theta) +C.F.
\end{align}
Collected, and using switching parameter $s$, we have:
\begin{equation}\label{paintedseries}
y_2=A\sin(t+\theta)-\epsilon s\frac{A}{2}t\sin(t+\theta) +\epsilon^2 s^2\frac{A}{8}\left[t^2\sin(t+\theta)-t\cos(t+\theta)\right].
\end{equation}
The determining equation for the generator to second order is:
\begin{equation}
\bm{X}_2(y-y_2)|_{y=y_2}=0
\end{equation}
Solving order-by-order, at zeroth order we have:
\begin{equation}
(\eta^{(0)}-A\cos(t+\theta)\xi^{(0)})|_{y=A\sin(t+\theta)}=0.
\end{equation}
Requiring this to be true for any boundary conditions gives $\eta^{(0)}=\xi^{(0)}=0$.

At first order:
\begin{align}
0&=\left(\bm{X}^{(1)}(y-y^{(0)})-\bm{X}^{(0)}y^{(1)}\right)\!\!\bigg|_{y=y^{(0)}}\nonumber\\
&\quad+\epsilon^{-1}\bm{X}^{(0)}(y-y^{(0)})|_{y=y_1}\\
&=\left(\eta^{(1)}-A\cos(t+\theta)\xi^{(1)}+\frac{A}{2}t\sin(t+\theta)\right)\!\!\bigg|_{y=y^{(0)}}\\
&\Rightarrow  \xi^{(1)}=0,\ \eta^{(1)}=-\frac{y}{2}t.
\end{align}
At second order:
\begin{align}
0&=\bm{X}^{(2)}(y-y^{(0)})|_{y=y^{(0)}}-\bm{X}^{(0)}y^{(2)}|_{y=y^{(0)}}\nonumber\\
&\quad-\bm{X}^{(1)}y^{(1)}|_{y=y^{(0)}}+\epsilon^{-2}\bm{X}^{(0)}(y-y^{(0)})|_{y=y_2}\nonumber\\
&\quad+\epsilon^{-1}\bm{X}^{(1)}(y-y^{(0)})|_{y=y_1}+\epsilon^{-1}\bm{X}^{(0)}(y-y^{(1)})|_{y=y_1}.
\end{align}
Since $\bm{X}^{(1)}$ has only a non-zero $y$-component, and $\bm{X}^{(0)}$ a $s$-component, this simplifies to:
\begin{align}
0&=-s\frac{A}{4}\left[t^2\sin(t+\theta)-t\cos(t+\theta)\right]|_{y=y^{(0)}}\nonumber\\
&\quad+\eta^{(2)}-A\cos(t+\theta)\xi^{(2)}\nonumber\\
&\quad+\epsilon^{-1}\left(-\frac{y}{2}t+\frac{y^{(0)}}{2}t\right)\!\!\bigg|_{y=y_1}\\
&=\eta^{(2)}+A\cos(t+\theta)\left(\frac{st}{4}-\xi^{(2)}\right)\\
&\Rightarrow \eta^{(2)}=0,\quad \xi^{(2)}=s\frac{t}{4}
\end{align}
So, overall the generator is:
\begin{equation}\label{generatordetermined}
\bm{X}_2=\frac{\partial}{\partial s}+\epsilon^2 s\frac{t}{4}\frac{\partial}{\partial t}-\epsilon\frac{t}{2}y \frac{\partial}{\partial y}.
\end{equation}
The FT equations are:
\begin{equation}
\frac{dt}{ds}=\epsilon^2 s\frac{t}{4},\quad \frac{dy}{ds}=-\epsilon y \frac{t}{2}.
\end{equation}
Integrating the first of these, we obtain:
\begin{equation}
t=\tilde{t}e^{\epsilon^2s^2/8},
\end{equation}
where $\tilde{t}=t(s=0)$. The second FT equation becomes:
\begin{align}
\frac{dy}{ds}&=-\epsilon y \tilde{t}\frac{e^{\epsilon^2s^2/8}}{2}+O(\epsilon^3)\\
&=-\epsilon y\frac{\tilde{t}}{2}+O(\epsilon^3).
\end{align}
Integrating gives:
\begin{equation}
y=\tilde{y}e^{-\epsilon s\tilde{t}/2}.
\end{equation}
Setting $s=1$ and substituting into the special $s=0$ solution $\tilde{y}=A\sin(\tilde{t}+\theta)$, yields the global solution:
\begin{equation}
y=e^{-\epsilon t/2}\sin(t e^{-\epsilon^2/8}+\theta)+O(\epsilon^3).
\end{equation}

Note, the symmetry generator Eq.~\eqref{generatordetermined} can also be calculated very laboriously using CAS by calculating first the approximate extended symmetries in $s,\ y$ and $t$ of the parent DE, and then restricting these on the perturbative solution. This could be viewed as an explicit verification of the above calculation.

\section{Supplemental figures}
\FloatBarrier
\begin{figure}
	\centering
	\includegraphics[width=0.48\textwidth]{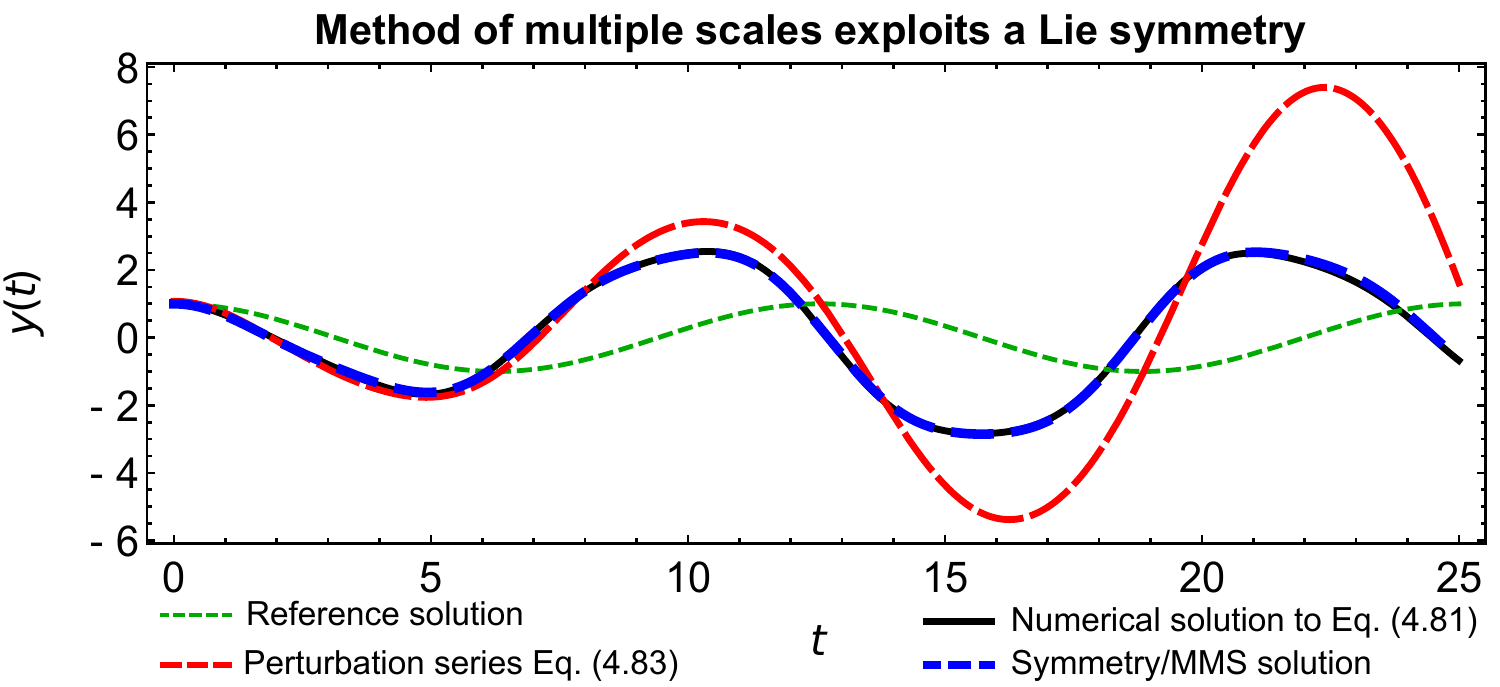}
	\caption{The hidden scale symmetry method is easily able to solve the Mathieu equation Eq.~\eqref{MathieuDE} (plotted with $\epsilon=0.15,\ a_1=1.2$) to first order, previously explored using MMS in ref.~\cite{Bender1999I} and CGO RG in ref.~\cite{Chen1996}. The selective painting procedure necessary to do so makes clear that MMS seeks a hidden scale symmetry connecting slow timescales and integration constants.}
	\label{FigS1}
\end{figure}
\FloatBarrier

\end{document}